\documentclass[twocolumn]{aastex701}
\usepackage{amsmath} 
\usepackage{tabularx,booktabs} 
\usepackage{gensymb} 
\usepackage{subcaption} 

\usepackage{xcolor}
\usepackage{float}  
\usepackage{braket}

\begin{document}

\title{Accretion-Driven Evolution of Compact-Object Populations in Gas-Rich Environments and the Origin of Massive Gravitational-Wave Sources}

\author[orcid=0000-0002-2728-0132,sname='Rozner']{Mor Rozner} 
\affiliation{Institute for Advanced Study, Einstein Drive, Princeton, NJ 08540, USA}
\affiliation{Institute of Astronomy, University of Cambridge, Madingley Road, Cambridge CB3 0HA, UK}
\affiliation{Gonville \& Caius College, Trinity Street, Cambridge CB2 1TA, UK}
\email[show]{morozner@ias.edu}

\author[orcid=0000-0002-9537-1933,sname='Rosselli-Calderon']{Alejandra Rosselli-Calderon} 
\affiliation{Department of Astronomy and Astrophysics, University of California, Santa Cruz, CA 95064, USA}
\email{aleroca@ucsc.edu}

\author[orcid=0000-0003-2558-3102,sname='Ramriez-Ruiz']{Enrico Ramirez-Ruiz} 
\affiliation{Department of Astronomy and Astrophysics, University of California, Santa Cruz, CA 95064, USA}
\email{enrico@ucolick.org}

\begin{abstract}
The origin of the most massive gravitational-wave sources remains elusive. We show that gas accretion can be understood as a transport process in mass space, causing compact objects to migrate through a population at rates determined by the underlying growth law. Using a continuity-equation framework, we demonstrate that population evolution is governed primarily by the mass dependence of the accretion rate, $\dot m \propto m^\beta$. Accretion laws with $\beta>1$ naturally produce divergent evolution and generate extended high-mass tails, whereas $\beta<1$ leads to convergent evolution and compresses the population toward a narrower range of masses. We apply this framework to physically motivated accretion regimes and explore their consequences using analytical calculations and Monte Carlo population models. We show that sustained gas accretion can substantially broaden compact-object mass distributions, populate the high-mass end of gravitational-wave catalogs, and alter the mass-ratio distribution of compact-object binaries. In particular, collective accretion within compact binaries drives their mass ratios toward unity. Our results suggest that gaseous environments act as transport media that continuously reshape compact-object populations, providing a natural pathway toward the formation of massive mergers such as GW231123 and the high-mass tails increasingly revealed by gravitational-wave observations.
\end{abstract}

\section{Introduction}
The mass function is one of the most fundamental properties of an astrophysical population. Mass distributions encode information about formation processes, but they are not immutable fossils of their birth conditions. Subsequent evolution through stellar evolution, dynamical interactions, mergers, and accretion can substantially reshape a population. Understanding how these processes transform mass functions is therefore essential for connecting observed populations to their origins \citep[e.g.,][]{Zinnecker1982,Bonnell2001a,Bonnell2001b}.

Gas-rich environments offer a unique setting in which the evolution of individual compact objects and the evolution of entire populations become intimately linked. Through the exchange of mass, energy, and angular momentum with their surroundings, embedded objects can experience substantial growth and dynamical evolution over their lifetimes. While the evolution of individual accreting objects has been studied extensively, considerably less attention has been paid to how accretion reshapes entire populations. These environments are also expected to provide fertile grounds for compact-object interactions and mergers, making them particularly relevant in the era of gravitational-wave astronomy \citep[e.g.,][]{McKernan2012,Stone2017,Tagawa2020,Tagawa2021,RoznerPerets2024}.

Gas accretion not only increases the masses of individual objects; it changes the evolution of the population itself. Because accretion rates generally depend on mass, different members of a population evolve at different rates, inducing a systematic transport of the population through mass space. As a result, accretion can alter the shape of the mass function itself, broaden or narrow distributions, modify binary mass-ratio distributions, and ultimately change the observable properties of merger populations \citep[e.g.,][]{Zinnecker1982,Roupas2025}. These effects are expected to be particularly important for compact-object populations embedded in gas-rich environments, where accretion and dynamical interactions can proceed simultaneously over extended periods.

The rapidly expanding gravitational-wave catalog now provides an opportunity to probe these evolutionary effects directly. Since the first detection of gravitational waves \citep{FirstGW}, the number of observed mergers has increased dramatically, enabling increasingly precise studies of compact-object populations and their statistical properties. One of the most intriguing features revealed by gravitational-wave observations is the black-hole mass distribution, which exhibits a broad concentration of systems with primary masses of $\approx10$--$40\,M_\odot$, a relative scarcity of black holes between roughly $50$ and $120\,M_\odot$, and an extended high-mass tail whose origin remains uncertain \citep{RakavyShaviv1967,WoosleyHeger2021}.

Particularly striking examples, such as GW231123 \citep{GW231123detection}, with inferred component masses of $m_1 = 137^{+22}_{-17}\,M_\odot$ and $m_2 = 103^{+20}_{-52}\,M_\odot$, have renewed interest in mechanisms capable of populating the upper end of the black-hole merger distribution. Several pathways have been proposed to explain the formation of such systems, including hierarchical mergers in dense stellar environments \citep[e.g.,][]{Fishbach2017,Yang2019,Antonini2025}, gas-assisted growth through prolonged accretion \citep[e.g.,][]{BartosHaiman2025,Roupas2025,Tagawa2026}, and combinations of both channels \citep[e.g.,][]{Kiroglu2025,Tagawa2026}. Each pathway predicts a different evolutionary history and therefore leaves distinct imprints on the observed black-hole population. Rather than explaining a single gravitational-wave event, we develop a general framework for understanding how gas accretion reshapes compact-object populations and the merger distributions they produce.

In this Letter, we develop a general framework for understanding how gas accretion transforms compact-object populations. We demonstrate that accretion not only grows individual black holes but also induces a systematic transport through mass space that reshapes mass functions, modifies binary properties, and alters the distribution of gravitational-wave mergers. We connect this framework to physically motivated accretion prescriptions, explore the evolution of compact-object binaries and their mass-ratio distributions, and investigate the resulting signatures in gravitational-wave populations. Together, these results establish a direct connection between the statistical properties of compact-object populations and the gaseous environments through which they evolve.

\section{Accretion-Driven Transport in Mass Space}\label{sec:distributions}

The evolution of an accreting population can be viewed as a transport problem in mass space, analyzed using the continuity equation. As compact objects accrete gas, they move toward larger masses at rates determined by the underlying growth law. The resulting evolution of the population is governed by a continuity equation. In this picture, accretion does not simply increase the masses of individual objects; it drives a collective redistribution of the whole population across mass space.

Let $p_0(m_0)$ be the initial mass function. We consider a population of compact objects that evolve from an initial mass $m_0$ to a mass $m$ after a time $t$. Conservation of objects requires that systems are neither created nor destroyed as they move through mass space. The number of objects contained within a mass interval must therefore remain constant along the evolution, implying
\begin{equation}
p(m,t)\,dm = p_0(m_0)\,dm_0,
\end{equation}

or equivalently
\begin{equation}
p(m,t)=p_0(m_0)\left|\frac{dm_0}{dm}\right|,
\end{equation}
where the Jacobian $\left|dm_0/dm\right|$ is determined by the underlying growth law. This Lagrangian formulation is equivalent to the continuity equation and provides a convenient framework for following the evolution of mass functions under accretion \citep{Zinnecker1982}. Sink terms associated with mergers, or source terms associated with compact-object formation, can be incorporated straightforwardly, but are neglected here in order to isolate the effects of accretion-driven transport.

To develop a general description of population evolution, we consider accretion laws of the form
\begin{equation}
\dot m = k m^\beta e^{-t/\tau_{\rm gas}},
\label{eq:mdot}
\end{equation}
where $k$ is a normalization constant, $\beta$ determines how the accretion rate scales with mass, and $\tau_{\rm gas}$ characterizes the lifetime of the gas reservoir. This form encompasses a broad class of physically relevant accretion prescriptions and allows the evolution of the mass function to be studied independently of any particular astrophysical environment.

The parameter $\beta$ is the key quantity governing the evolution of the population. In this framework, 
the qualitative behaviour of the population is determined  primarily by the value of $\beta$, which determines how rapidly different regions of the mass function evolve and therefore whether accretion tends to broaden, preserve, or compress the distribution.    The overall factor $k$ could encapsulate details of the astrophysical environment that determine the accretion efficiency and the final masses.

For $\beta<1$, lower-mass objects grow relatively faster than higher-mass objects, leading to convergent evolution and a narrowing of the mass distribution. For $\beta=1$, all objects experience the same fractional growth rate and the mass function evolves self-similarly, preserving its overall shape while shifting toward higher masses. In contrast, for $\beta>1$, more massive objects grow disproportionately rapidly, producing divergent evolution, expanding the original distribution, and naturally generating extended high-mass tails.

For a broad class of initial mass functions, analytical solutions can be obtained when the accretion rate follows Equation~\ref{eq:mdot}. These solutions make explicit how the growth law governs the transport of objects through mass space and the resulting evolution of the population. As an illustrative example, we consider an initial power-law distribution, $p_0(m_0)\propto m_0^\alpha$ with $\alpha<0$, and an accretion law $\dot m = km^\beta e^{-t/\tau_{\rm{gas}}}$, which allows the transport induced by accretion to be characterized analytically. The evolution of an object's mass is then given by

\small{
\begin{align}
&m(t)= \\ \nonumber =
&\begin{cases}
\left[m_0^{1-\beta}
+k\tau_{\rm gas}(1-\beta)
\left(1-e^{-t/\tau_{\rm gas}}\right)
\right]^{1/(1-\beta)},
& \beta\neq1,\\
m_0
\exp\left[
k\tau_{\rm gas}
\left(1-e^{-t/\tau_{\rm gas}}\right)
\right],
& \beta=1.
\end{cases}
\end{align}}
\normalsize

At late times ($t\rightarrow\infty$), 
\begin{align}
m_\infty=
\begin{cases}
\left[
m_0^{1-\beta}
+k\tau_{\rm gas}(1-\beta)
\right]^{1/(1-\beta)},
& \beta\neq1,\\
m_0e^{k\tau_{\rm gas}},
& \beta=1.
\end{cases}
\end{align}

The corresponding asymptotic mass function becomes
\small{\begin{align}
p_\infty(m)\propto
\begin{cases}
\left[
m^{1-\beta}
-k\tau_{\rm gas}(1-\beta)
\right]^{(\alpha+\beta)/(1-\beta)}
m^{-\beta},
& \beta\neq1,\\
m^\alpha
e^{-k\tau_{\rm gas}(\alpha+1)},
& \beta=1.
\end{cases}
\label{eq:pm}
\end{align}}
\normalsize
Equation~\ref{eq:pm} shows explicitly that accretion modifies the mass function through the differential transport of objects across mass space. Because the transport velocity term in the continuity equation depends on mass, different regions of the distribution evolve at different rates.  The resulting mass function therefore reflects both its initial conditions and the underlying accretion physics.

The asymptotic form of the distribution is determined by the interplay between the initial mass function and the mass dependence of the accretion rate. Growth laws with $\beta>1$ generate extended high-mass tails, whereas $\beta<1$ compresses the population toward a narrower mass distribution.

Different accretion prescriptions reshape populations in distinct ways and consequently leave characteristic signatures on the resulting mass function. Observable features such as high-mass tails, broadened mass distributions, modified binary populations, and ultimately the properties of gravitational-wave merger populations can therefore be interpreted as the cumulative imprint of accretion-driven evolution. In the next section, we connect this general framework to the physical accretion regimes expected in gas-rich environments.

\section{Accretion-Driven Growth Laws}\label{sec:acc}

The framework developed in Section~\ref{sec:distributions} shows that the evolution of an accreting population is governed primarily by the mass dependence of the growth law, parameterized by $\beta$ and $k$. We now connect this general description to the physical accretion regimes expected in gas-rich environments. Different functional forms of accretion prescriptions correspond to different values of $\beta$ and therefore drive distinct patterns of population evolution.

The gas accretion rate onto an object of mass $m$ is commonly written as
\begin{equation}
\dot m=\pi \rho_g v_{\rm eff}R_{\rm acc}^2,
\end{equation}
where $\rho_g$ is the gas density, $v_{\rm eff}$ is the effective relative velocity between the gas and the accretor, and $R_{\rm acc}$ is the characteristic accretion radius. Different physical assumptions regarding $R_{\rm acc}$ lead to different mass scalings and therefore different values of $\beta$.

\subsection{Bondi-Hoyle-Lyttleton Accretion ($\beta=2$)}

In the Bondi-Hoyle-Lyttleton (BHL) regime, the accretor is treated as moving through an effectively unbounded gaseous medium \citep{Bondi1952,HoyleLyttleton1941}. The characteristic accretion radius is
\begin{equation}
R_{\rm BHL}=\frac{2Gm}{v_{\rm eff}^2},
\end{equation}
where $v_{\rm eff}^2=v_{\rm rel}^2+c_s^2$ includes the relative velocity between the object and the gas with the sound speed. Since $R_{\rm BHL}\propto m$, the resulting accretion rate scales as
\begin{equation}
\dot m \propto m^2,
\end{equation}
corresponding to $\beta=2$. As discussed in Section~\ref{sec:distributions}, such growth laws produce divergent evolution, naturally broadening the mass distribution and enhancing its high-mass tail.

\subsection{Hill Accretion ($\beta=2/3$)}

In environments where accretion is influenced by a dominant central potential, the relevant cross-section for accretion is determined by the Hill radius,
\begin{equation}
R_{\rm H}=r\left(\frac{m}{3M_{\rm SMBH}}\right)^{1/3},
\end{equation}
where $M_{\rm SMBH}$ is the mass of the central supermassive black hole and $r$ is the orbital radius \citep{Bonnell2001a,Bonnell2001b,2023ApJ...944...44K}. Since $R_{\rm Hill}\propto m^{1/3}$, the accretion rate scales as
\begin{equation}
\dot m \propto m^{2/3},
\end{equation}
corresponding to $\beta=2/3$. In contrast to BHL accretion, this regime produces convergent evolution and tends to suppress differential growth across the population.

\subsection{Eddington-Limited Growth ($\beta=1$)}

Accretion is conservatively ultimately limited by radiation feedback. The Eddington luminosity is
\begin{equation}
L_{\rm Edd}
=\frac{4\pi G m m_p c}{\sigma_T},
\end{equation}
which corresponds to an Eddington-limited accretion rate
\begin{equation}
\dot m_{\rm Edd}
=
\frac{4\pi G m m_p}
{\epsilon \sigma_T c},
\end{equation}

where $m_p$ is the proton mass, $\epsilon$ is the radiative efficiency, and $\sigma_T$ is the Thomson cross section. Since
\begin{equation}
\dot m_{\rm Edd}\propto m,
\end{equation}
this corresponds to $\beta=1$, for which all objects experience the same fractional growth rate and the mass function evolves self-similarly.

It should be noted that limiting the accretion by the Eddington limit is somewhat conservative, as super-Eddington accretion might take place under certain conditions \citep{Abramowicz1988}. 

Despite their different physical origins, these accretion regimes all map naturally onto the transport framework developed in Section~\ref{sec:distributions}, and the differences between them are encapsulated in different Jacobians, i.e. different $|dm_0/dm|$. Their primary distinction lies in the value of $\beta$, which determines whether the resulting population evolution is convergent, self-similar, or divergent.

\begin{figure}[h]
\centering
\includegraphics[width=1\linewidth]{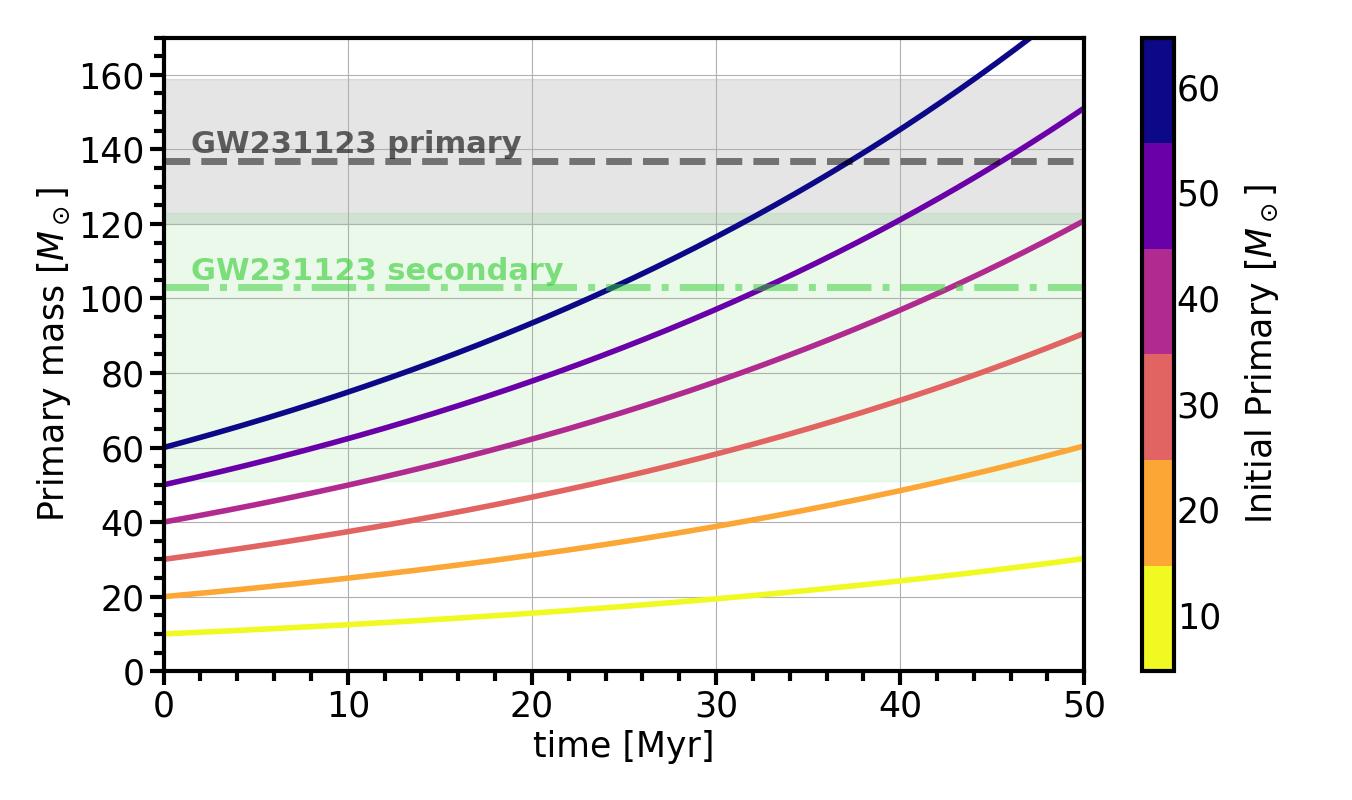}
\caption{
Illustration of accretion-driven transport in mass space for the accretion regimes described above. Different colours show the evolution of compact objects with different initial masses under the fiducial gas-rich conditions adopted in this work. The calculations assume a local gas density of $7\times10^{-16}\,{\rm g\,cm^{-3}}$ and a gas lifetime of $\tau_{\rm gas}=10\,{\rm Myr}$, representative of conditions at $r=0.1\,{\rm pc}$ in an AGN disk surrounding a $10^7\,M_\odot$ supermassive black hole. Because the accretion rate remains mass dependent, more massive objects evolve more rapidly than lower-mass objects, providing a concrete example of the differential growth discussed in Section~\ref{sec:distributions}. The dashed and dot-dashed horizontal lines indicate the primary and secondary masses inferred for GW231123, while the shaded regions denote the corresponding uncertainties.
}
\label{fig:traj}
\end{figure}

Figure~\ref{fig:traj} illustrates the trajectories of compact objects in a representative AGN-disk environment, where growth is regulated by the Hill accretion prescription. Even in this convergent regime ($\beta=2/3$), more massive objects evolve more rapidly than lower-mass objects, providing a concrete example of the differential growth discussed in Section~\ref{sec:distributions}. These trajectories foreshadow how accretion-driven transport reshapes the compact-object populations presented in the following section.

In the next section, we apply these growth laws to investigate how accretion-driven transport reshapes compact-object populations and modifies their observable mass distributions.

\section{Evolution of Compact-Object Binary Populations}\label{sec:bindis}

While the transport framework developed above describes the evolution of individual accretors, compact objects frequently reside in binaries. In addition to increasing the total mass of a binary, gas accretion can redistribute mass between the two components and therefore modify the binary mass ratio. Understanding how gas is captured and partitioned within binaries is therefore essential for predicting the evolution of compact-object populations.

When the binary separation greatly exceeds the characteristic accretion radius of either component, each object accretes approximately independently \citep[e.g.,][]{2019ApJ...876..142K}. In this limit, the binary can be treated as two isolated accretors whose masses evolve according to the growth laws discussed in Section~\ref{sec:acc}.

As the ratio $a_{\rm bin}/R_{\rm acc}$ decreases, the gravitational spheres of influence of the two objects begin to overlap. The gas can no longer be regarded as responding independently to each component. Instead, the binary increasingly interacts with the surrounding medium through its combined gravitational potential, leading to the formation of a common wake and enhanced gravitational focusing \citep{Antoni2019}.

For binaries with separations $a_{\rm bin}\lesssim R_{\rm acc}$, the companions share a common accretion flow and interact with the surrounding gas as a single gravitational system \citep{Antoni2019}. In this regime, the total accretion rate is determined by the combined mass of the binary rather than the masses of the individual components. In the BHL limit, this implies
\begin{equation}
\dot M \propto (m_1+m_2)^2,
\end{equation}
which generally exceeds the combined accretion rates expected for two isolated accretors \citep{Antoni2019}.

The evolution of a binary is governed not only by the total amount of gas captured by the system but also by how that gas is distributed between the two components. Hydrodynamic simulations show that once a common wake develops, the secondary experiences an enhanced local gas density relative to the ambient medium \citep{Rosselli-Calderon2026}. Consequently, the lower-mass companion can accrete more efficiently than it would in isolation, causing the binary mass ratio to evolve toward unity.

Following \citet{Antoni2019}, we compute the total accretion rate using the common-accretion prescription and incorporate the density enhancement measured by \citet{Rosselli-Calderon2026}, which scales approximately as $(a_{\rm bin}/R_{\rm acc})^{-1}$. Alternative prescriptions involving mini-disks and different mass-partition schemes have also been proposed \citep{Gerosa2015,Comerford2019,YoungClarke2015}. Although they differ in the details of the accretion flow, they generally predict qualitatively similar mass-ratio evolution.

For the calculations presented below, we adopt a sub-Keplerian relative velocity between the binary and the surrounding gas, $\eta_{\rm gas}v_{\rm Kep}$, where $\eta_{\rm gas}\approx1.5(c_s/v_{\rm Kep})^2$ is the pressure-support parameter of the disk and $v_{\rm Kep}$ is the local Keplerian velocity. These assumptions fully specify the binary growth model. We now examine how the resulting accretion-driven transport reshapes the mass distribution of compact-object populations before turning to its impact on binary mass ratios.

\subsection{Population Evolution}

We perform Monte Carlo simulations to follow the evolution of accreting compact-object binaries and the populations they comprise. The transport framework developed in Section~\ref{sec:distributions} predicts that accretion reshapes a population by driving objects through mass space at rates that depend on their masses. We now test these predictions using population models initialized with physically motivated black-hole mass distributions.

The black-hole mass function in gas-rich environments remains uncertain and may reflect a combination of stellar evolution, dynamical processing, and previous accretion episodes \citep[e.g.,][]{Sukhbold2016,Yang2019a,Patton2022}. To illustrate the effects of accretion-driven transport, we adopt an initial distribution consisting of a Salpeter-like power law \citep{Salpeter1955} combined with a Gaussian component,
\begin{equation}
\xi(m)=C_1m^{-2.3}+C_2
\exp\left[-\frac{(m-\mu)^2}{2\sigma^2}\right],
\end{equation}
with $\mu=10\,M_\odot$ and masses drawn in the range $5-50\,M_\odot$.

\begin{figure}
    \centering \includegraphics[width=1\linewidth]
       {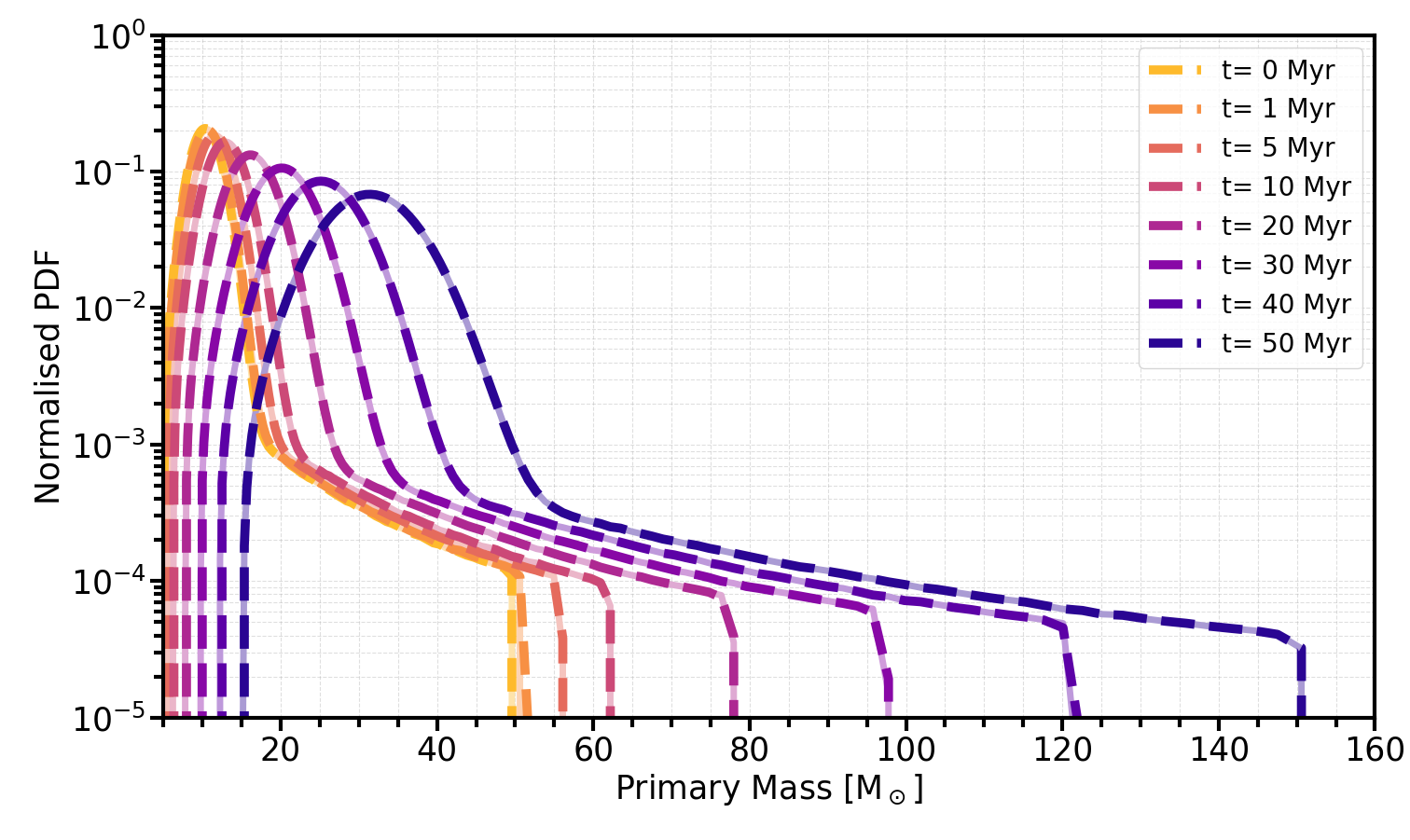}
       \caption{
Evolution of the normalized primary-mass distribution under sustained gas accretion. The initial population consists of a Salpeter-like power law combined with a Gaussian component centered at $\mu=10\,M_\odot$, with masses drawn between $5$ and $50\,M_\odot$. The gas density evolves as $\rho_g(t)=7\times10^{-16}\exp(-t/\tau_{\rm gas})\,{\rm g\,cm^{-3}}$ with $\tau_{\rm gas}=10\,{\rm Myr}$. Different colors show the mass function at successive times. Solid curves correspond to Monte Carlo calculations and dashed curves to the analytical solutions derived using the continuity equation framework. The distribution broadens and extends toward higher masses as accretion drives objects through mass space at mass-dependent rates.
}
    \label{fig:Normalised_mass_function}
\end{figure}

Figure~\ref{fig:Normalised_mass_function} shows the evolution of the primary-mass distribution under sustained gas accretion. The solid curves correspond to the Monte Carlo calculations, and the dashed curves show the analytical solutions derived in Section~\ref{sec:distributions}. The close agreement between the two demonstrates that the evolution is governed by the transport framework developed above.

As accretion proceeds, the mass function broadens, the maximum mass increases, and the initial peak becomes progressively smeared. These trends arise because objects move through mass space at rates that depend on their mass. The resulting evolution is therefore more complicated than a uniform shift toward higher masses, and induces a continuous reshaping of the distribution itself.

Despite their different physical origins, these accretion regimes all map naturally onto the transport framework shown in Figure~\ref{fig:Normalised_mass_function}.  Massive objects migrate through mass space more rapidly than lower-mass objects, causing the distribution to evolve away from its initial form. Even in the Eddington-limited regime relevant for the adopted gas density, accretion substantially modifies the mass function on timescales of millions of years.

\begin{figure*}
    \centering
    \includegraphics[width=0.49\linewidth]{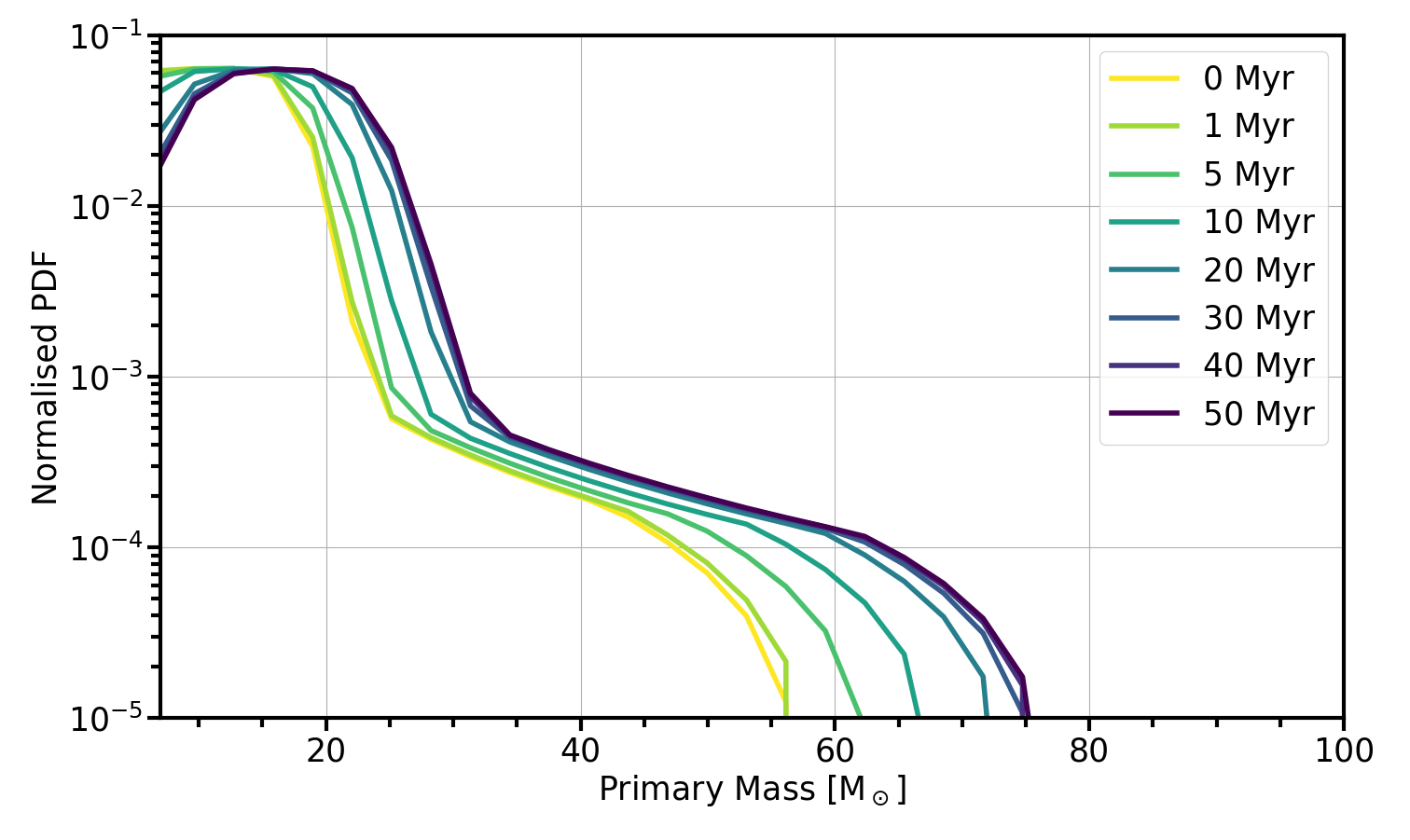}
\includegraphics[width=0.49\linewidth]{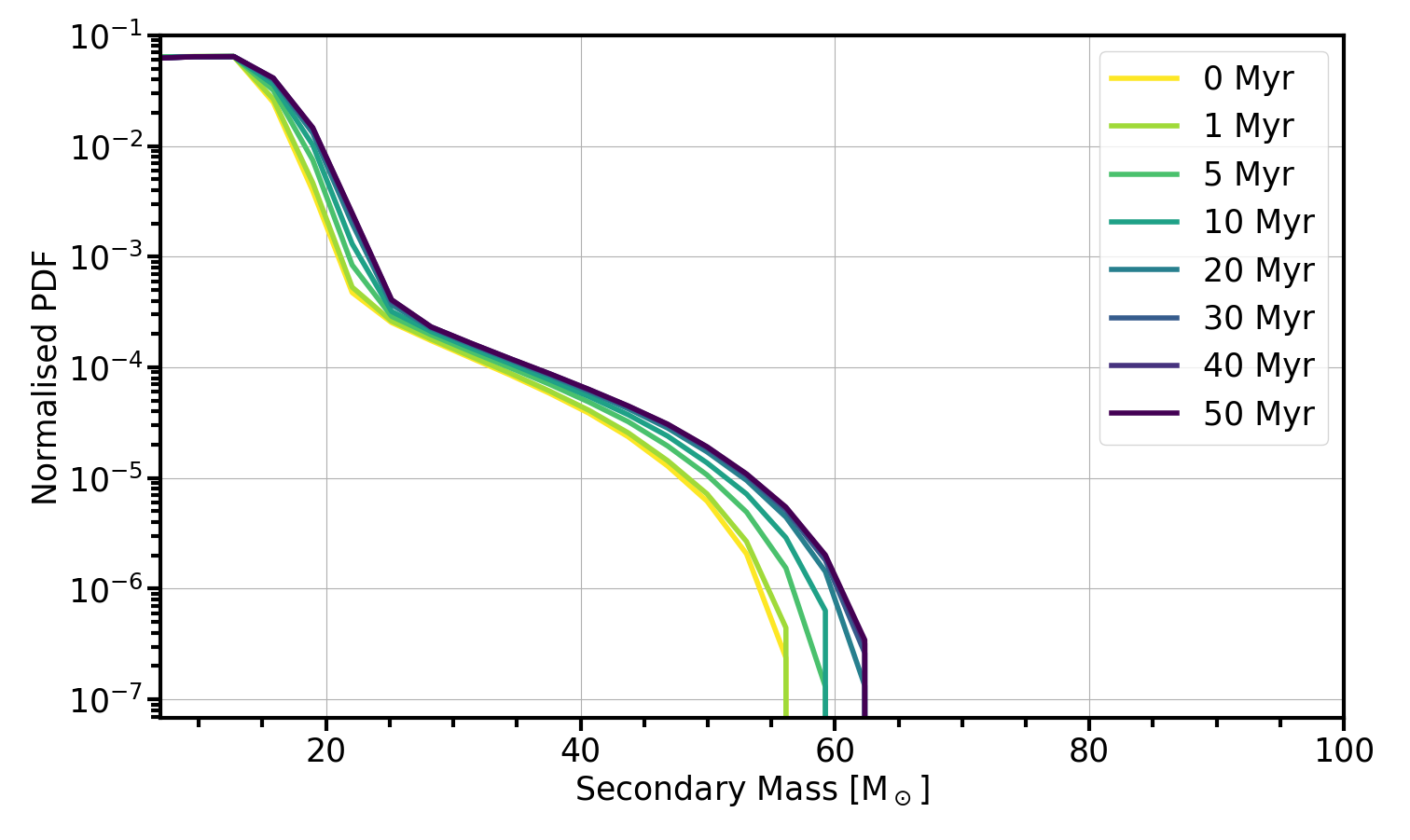}
\includegraphics[width=0.48\linewidth]{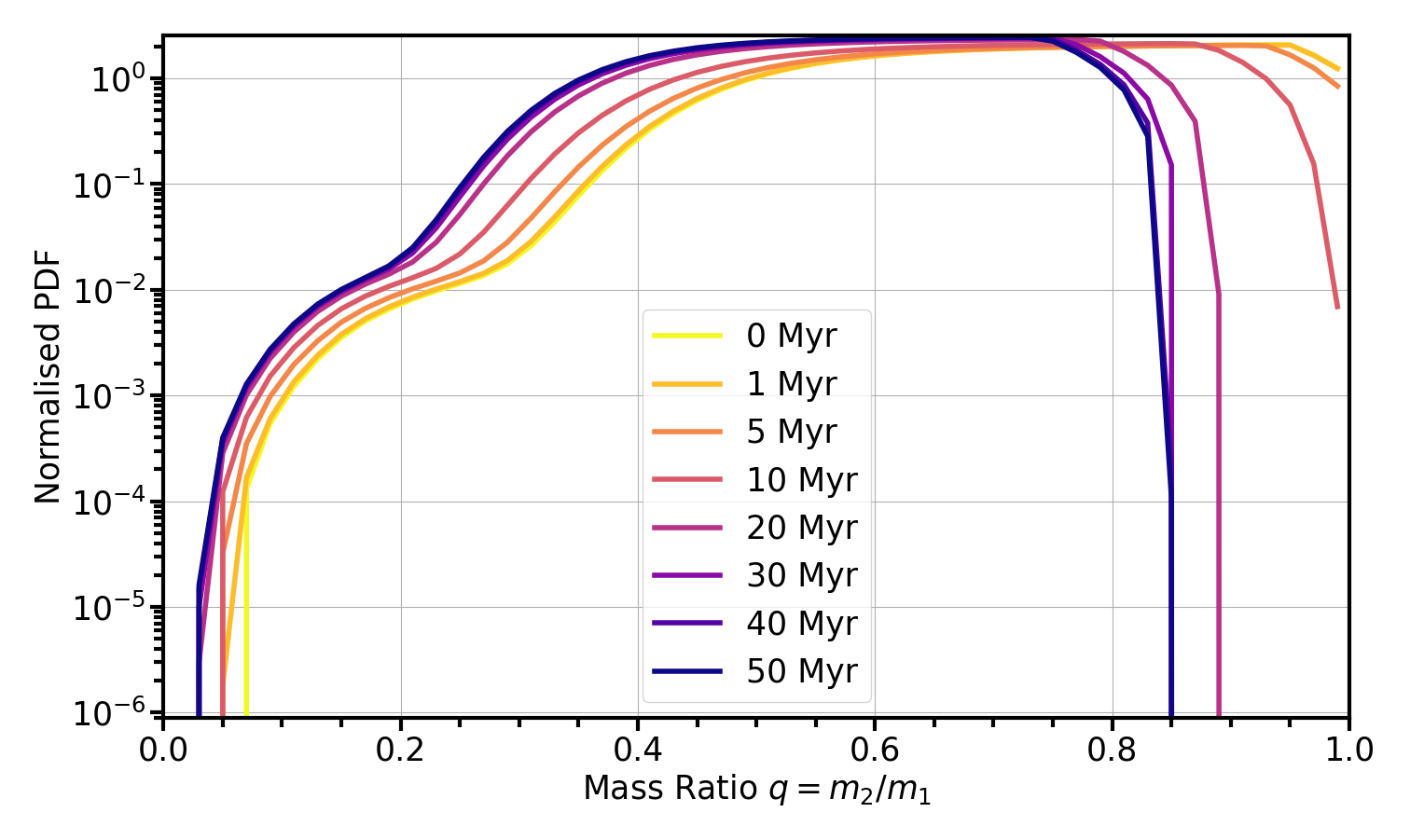}
    \caption{Time evolution of binary populations in the absence of secondary accretion enhancement, i.e. $\dot m_2= q^2/(1+q)^2\dot M, \ \dot m_1=(1+2q)/(1+q)^2$, $m_2$ accretes as isolated and $m_1$ completes to the total expected common accretion rate, according to the relevant regime.  
    The calculations are based on Monte Carlo simulations with $5\times10^6$ realizations, assuming an initial gas density of $\rho_{g,0}=10^{-18}{\rm{g \ cm^{-3}}}$ and a binary separation of $0.1 \ {\rm AU}$. The panels show the evolution of the primary-mass distribution (top left), secondary-mass distribution (top right), and mass-ratio distribution (bottom). Although both components grow through gas accretion, the more massive object accretes preferentially, causing the mass-ratio distribution to shift toward smaller values of $q$. In this limit, accretion amplifies the initial mass asymmetry between binary companions.
    }
    \label{fig:not_enhanced}
\end{figure*}

\begin{figure*}
    \centering
    \includegraphics[width=0.49\linewidth]{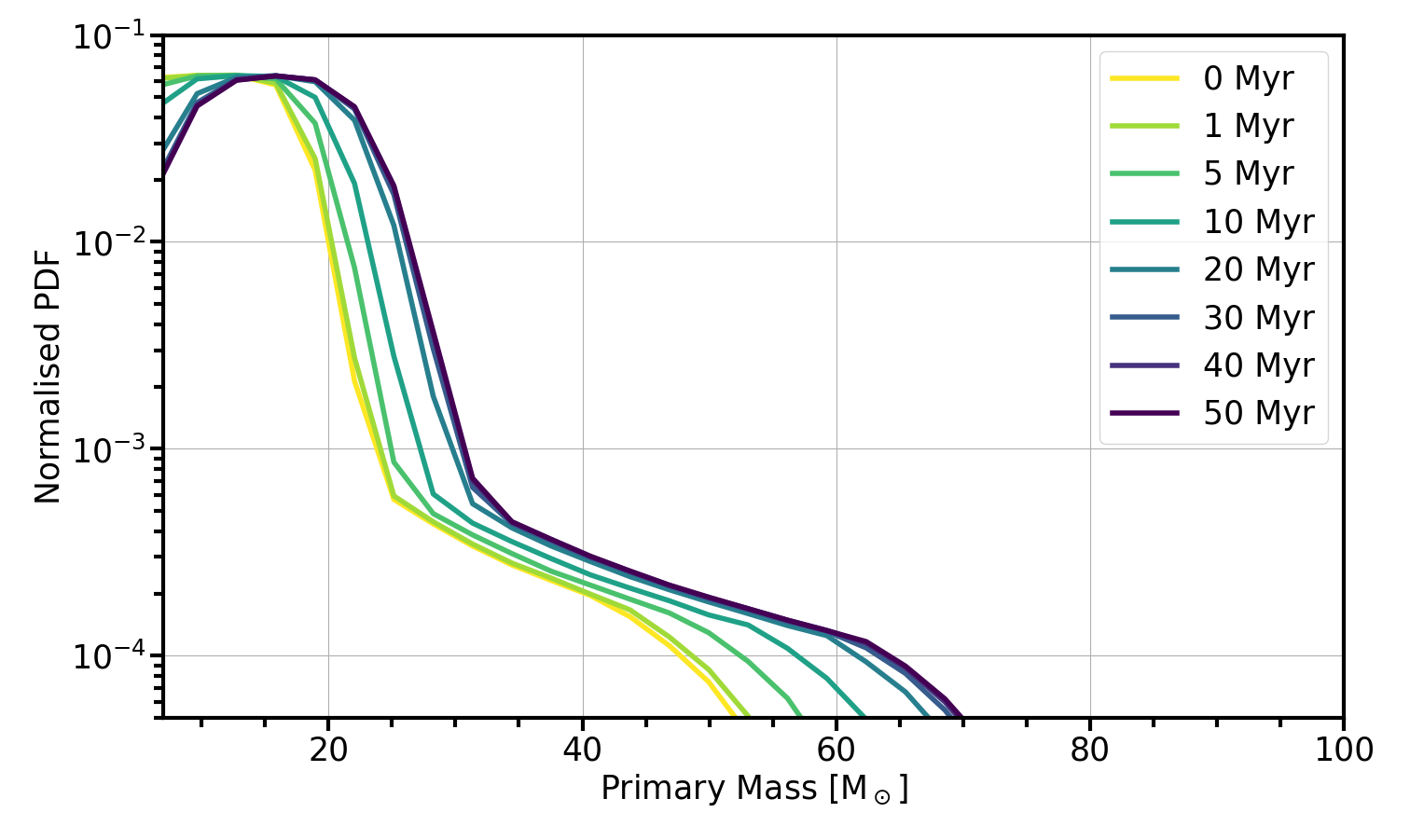}
    \includegraphics[width=0.49\linewidth]{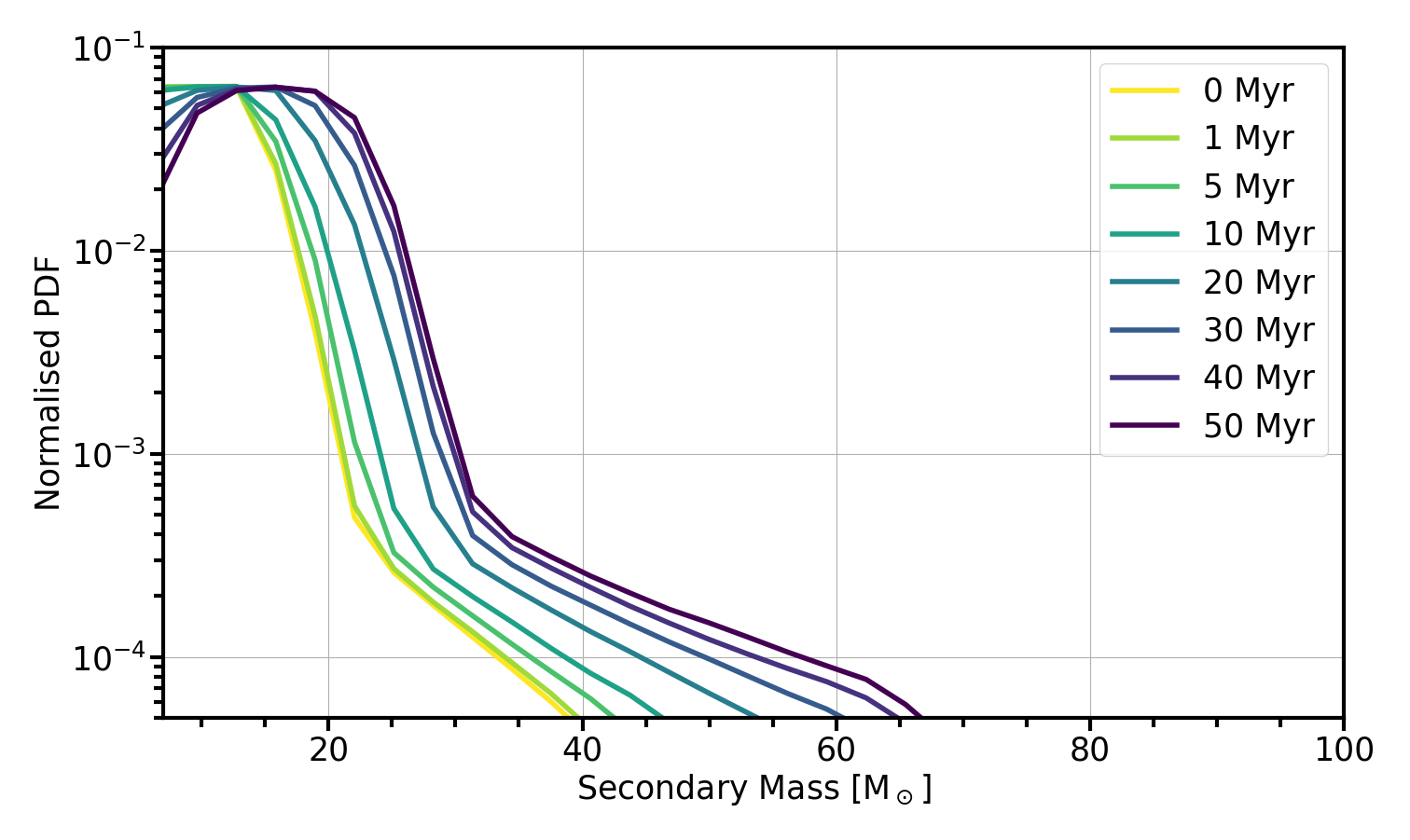}
\includegraphics[width=0.49\linewidth]{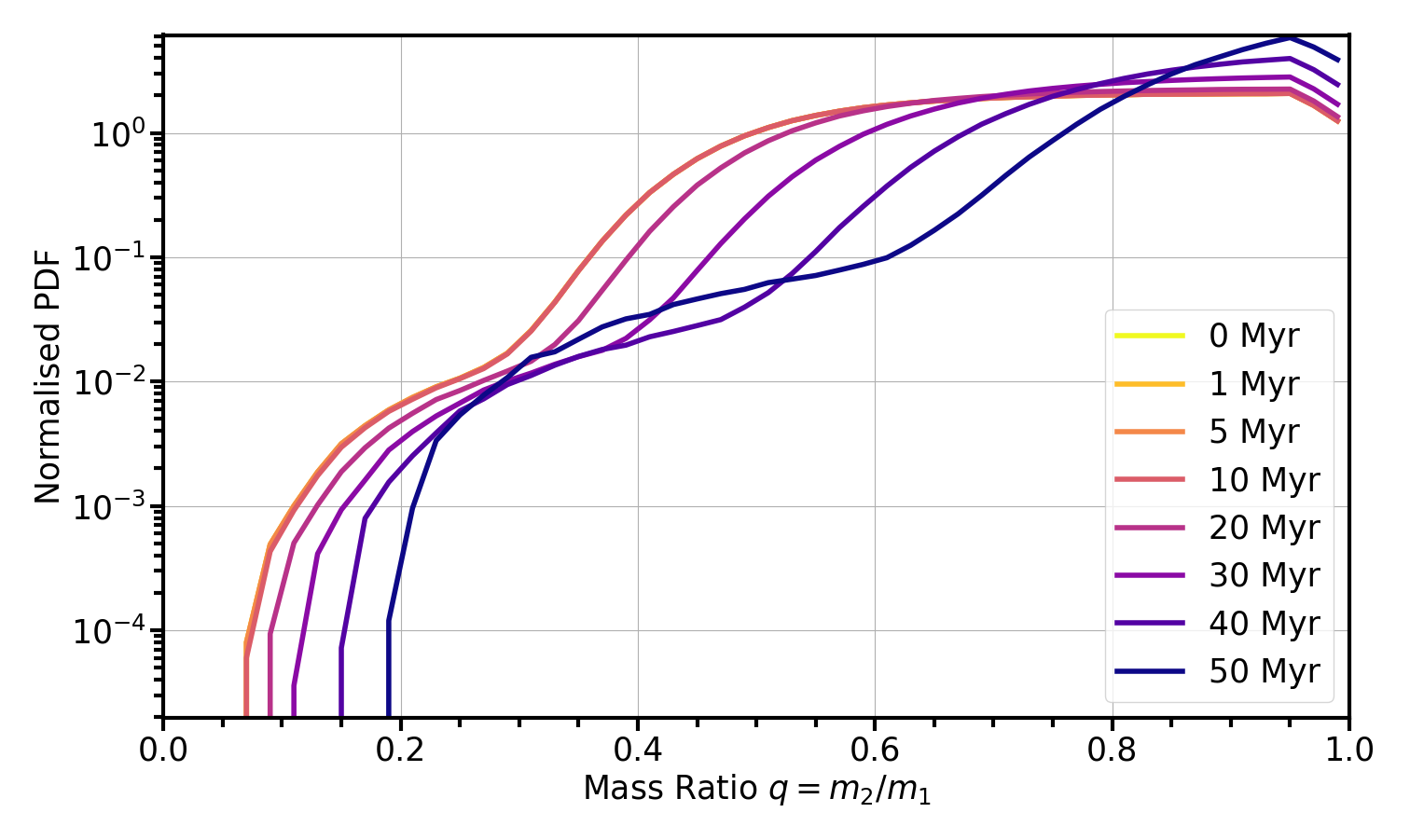}
    \caption{Time evolution of binary populations including enhanced accretion onto the secondary companion. The calculations are identical to those shown in Figure~\ref{fig:not_enhanced}, except that the secondary experiences the density enhancement associated with the common-accretion regime following \citet{Rosselli-Calderon2026}. The panels show the evolution of the primary-mass distribution (top left), secondary-mass distribution (top right), and mass-ratio distribution (bottom). In contrast to the isolated-accretion case, the enhanced growth of the secondary drives the mass-ratio distribution toward $q\simeq1$, causing the binary population to evolve toward increasingly equal-mass systems. The comparison with Figure~\ref{fig:not_enhanced} demonstrates that the partitioning of accreted gas between the binary components plays a crucial role in determining the properties of the resulting population.}
    \label{fig:enhanced}
\end{figure*}

\subsection{Mass-Ratio Evolution}
While the evolution of the primary-mass distribution reflects the transport of objects through mass space, binaries provide an additional observable through the evolution of their mass ratios. In gas-rich environments, the mass-ratio distribution is determined not only by the total amount of gas captured by a binary but also by how that gas is partitioned between its components. Different prescriptions for distributing the accreted material can therefore yield qualitatively distinct binary populations.

Defining the mass ratio as
\begin{equation}
q \equiv \frac{m_2}{m_1},
\end{equation}
its evolution is then governed by
\begin{equation}
\frac{dq}{dt}=
\frac{1}{m_1}
\left(
\frac{dm_2}{dt}-
q\frac{dm_1}{dt}
\right).
\end{equation}
If the two components accrete independently according to the same mass growth law,
\begin{equation}
\dot m = K m^\beta,
\end{equation}
then
\begin{equation}
\frac{dq}{dt}
=
K q m_1^{\beta-1}
\left(
q^{\beta-1}-1
\right).
\end{equation}
This expression shows that the evolution of the mass ratio is controlled directly by the mass dependence of the accretion rate. For $\beta>1$, the more massive component grows preferentially and the mass ratio decreases with time accordingly, driving binaries away from equal masses. For $\beta<1$, the lower-mass companion grows relatively more efficiently and the mass ratio evolves toward unity. The case $\beta=1$ corresponds to self-similar growth, for which the mass ratio remains unchanged.

These expectations apply when the binary components accrete independently from the surrounding medium. As discussed in the previous subsection, however, sufficiently compact binaries do not behave as two isolated accretors. Once the accretion radii overlap, the binary develops a common wake and interacts with the gas through its collective gravitational potential. The resulting redistribution of gas between the binary components can therefore produce mass-ratio evolution that differs substantially and qualitatively from the isolated-accretion limit.

Incorporating the enhancement of the secondary accretion rate associated with the common-accretion regime, the mass-ratio evolution becomes
\begin{align}
\frac{dq}{dt}=k m_1^{\beta-1}q\left(\frac{R_{\rm{acc}}}{a}q^{\beta-2}-1\right).
\end{align}

Unlike the isolated-accretion case, the evolution of the mass ratio now depends not only on the underlying accretion law but also on the degree of overlap between the binary and the common accretion flow. For sufficiently compact binaries, the enhanced growth of the secondary can overcome the preferential growth of the primary, driving the system toward equal masses. 

We now explore these effects using Monte Carlo population models. To explore these effects, we initialize binaries with a uniform mass-ratio distribution \citep[e.g.,][]{Bartos2017}, i.e., $p_0(q)=\mathrm{constant}$, subject to the requirement that the secondary mass exceeds $5\,M_\odot$. As a reference model, we first consider a scenario in which the secondary accretes according to its isolated accretion rate. Although the binary captures gas through its combined gravitational potential, the accreted material is partitioned according to the isolated growth rates of the two components.

Figure~\ref{fig:not_enhanced} shows the resulting evolution of the primary-mass, secondary-mass, and mass-ratio distributions. Because accretion rates increase with mass, the more massive component grows preferentially. The mass-ratio distribution therefore evolves toward smaller values of $q$, indicating that accretion deepens the differences between the binary components. In this limit, gas accretion drives binaries away from equal masses.

This behaviour changes qualitatively once the overlap of the accretion radii is taken into account. Hydrodynamic simulations indicate that the secondary resides within the overdense wake generated by the binary as a whole and therefore experiences a local gas density substantially larger than the ambient value \citep{Rosselli-Calderon2026}. We model this enhancement through
\begin{align}
\rho_{\rm{enh}}= \rho_\infty \left(\frac{m_1}{m_2}\right)^{\xi_1}\left(\frac{a}{R_{\rm{a}}}\right)^{-\xi_2}
\end{align}
where $\xi_1\approx\xi_2\approx1$. This prescription captures the tendency of the common wake to channel gas toward the lower mass companion once the binary enters the common accretion regime.

Figure~\ref{fig:enhanced} illustrates the opposite behaviour once collective accretion is taken into account. The enhanced growth of the secondary drives the mass-ratio distribution toward $q\simeq1$, demonstrating that common accretion naturally promotes the formation of nearly equal-mass binaries. The comparison between Figures~\ref{fig:not_enhanced} and \ref{fig:enhanced} demonstrates that the evolution of compact-object binaries depends not only on the total amount of gas accreted, but also on how that gas is distributed between the components. While isolated accretion tends to increase mass asymmetries, collective accretion within a common wake tends to drive binaries toward equal masses. 
Gas-rich environments, therefore, reshape not only the masses of compact objects but also the properties of the binaries that ultimately merge.

\section{Gravitational-Wave Populations}

The preceding sections demonstrate that gas accretion modifies both the mass distribution of compact objects and the mass-ratio distribution of compact-object binaries. These changes could be reflected directly in the properties of the gravitational-wave sources produced by gas-rich environments. In particular,
gas accretion can reshape the mass distribution and lead to a population of
 regions in parameter space that would otherwise be difficult to access through stellar evolution alone, while the tendency toward mass-ratio equalization can alter the distribution of merging binaries.

To quantify these effects, we calculate the merger-rate distribution as a function of primary mass,
\begin{align}
\Gamma_{\rm{merge}}&(t)=\\ \nonumber
&=N_{\rm{B}}n_{\rm{AGN}} \int \frac{p(m_1,t)p(m_2,t|m_1) p(a)}{t_{\rm{merge}}(m_1,m_2,a)} dm_1dm_2da
\end{align}
where $p(m_1,t)$ and $p(m_2,t|m_1)$ are the time-dependent primary and secondary mass distributions, $p(a)$ is the binary-separation distribution, $N_{\rm B}=100$ is the number of binaries, and $n_{\rm{AGN}}=4\times 10^4-3\times 10^6  \ \rm{Gpc}^{-3}$ \citep{Grobner2020} is the number density of AGN discs. 

We estimate the merger timescale using the standard gravitational-wave inspiral expression \citep{Peters1964}, 
\begin{equation}
t_{\rm merge}
= \left|
\frac{a}{\dot a_{\rm GW}}
\right|
\end{equation}
with
\begin{equation}
\dot a_{\rm GW}
=
-\frac{64G^3m_1m_2(m_1+m_2)}
{5c^5a^3}.
\end{equation}

For the separation distribution, we use \"{O}pik's law, i.e. $p(a)\propto a^{-1}$ \citep{Opik1924}, with separations in the range between $0.01 \ \rm{AU}$ and a maximum value of $10^5 \ \rm{AU}$, truncated at $10^2 \rm{AU}$ \citep[e.g.][]{Stone2017}. 
It should be noted that this is a conservative derivation of the rate, as the separation is expected to shrink in the presence of gas.

\begin{figure*}
    \centering
        \includegraphics[width=0.49\textwidth]
   {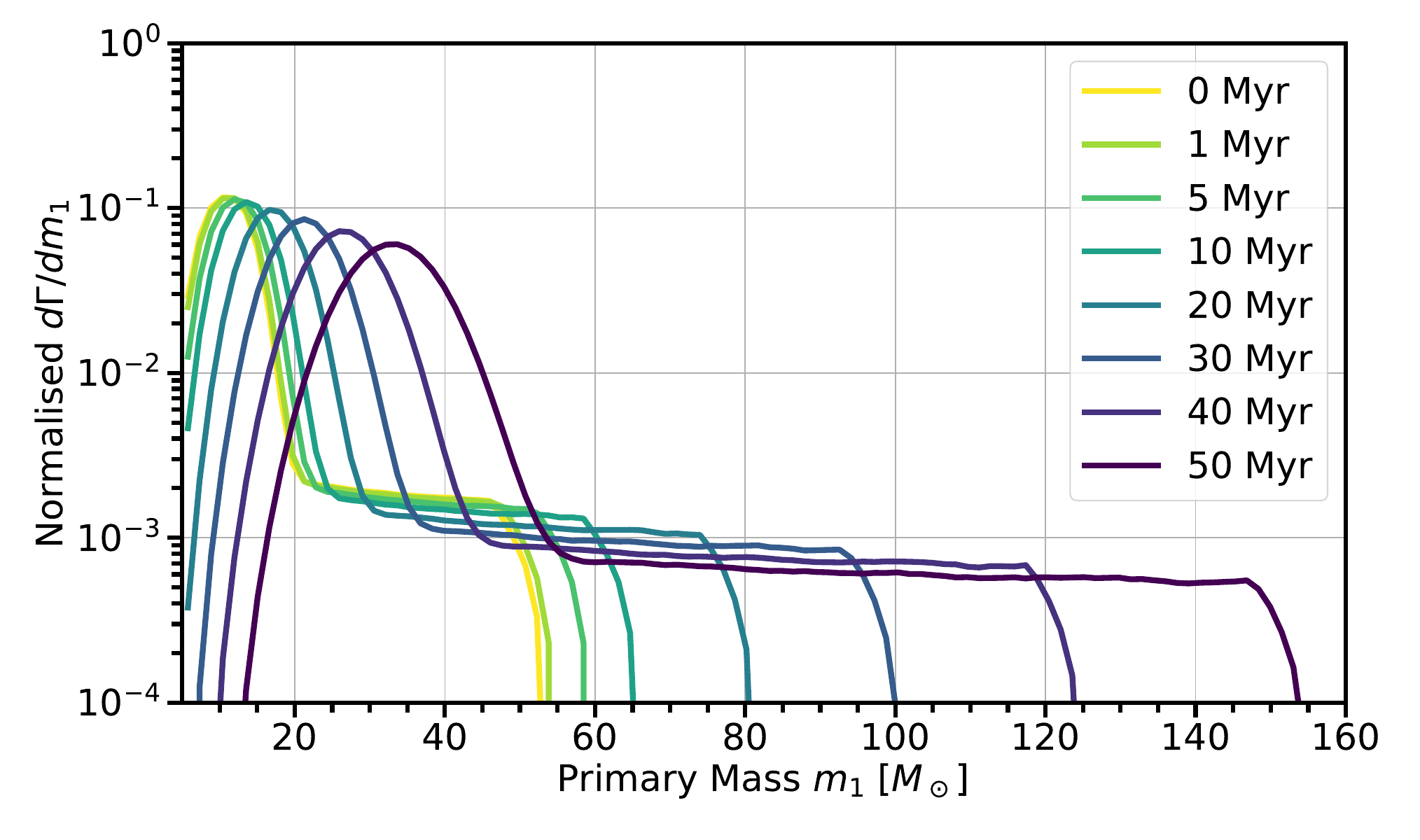}
    \includegraphics[width=0.49\textwidth]
    {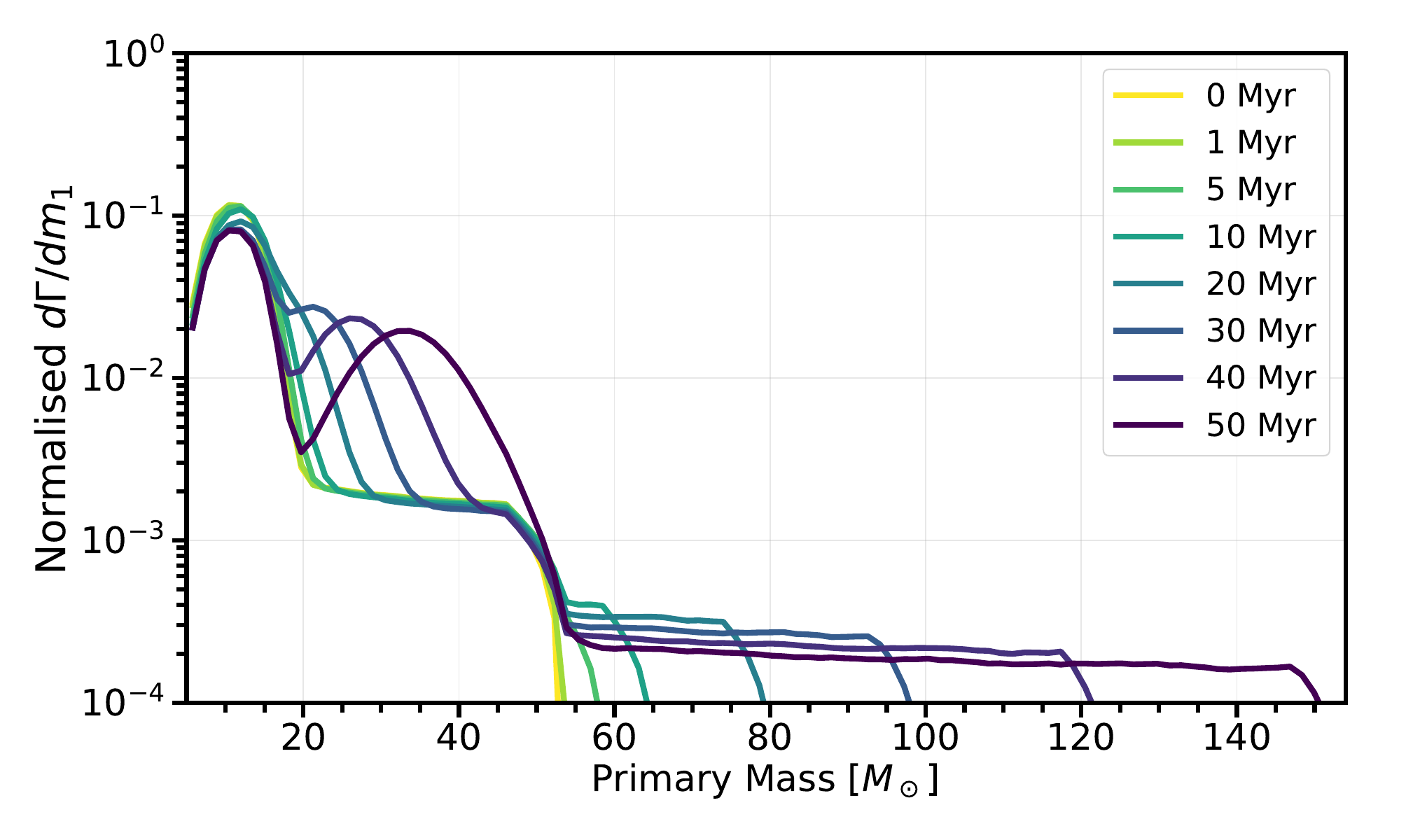}
\\\includegraphics[width=0.49\textwidth]
    {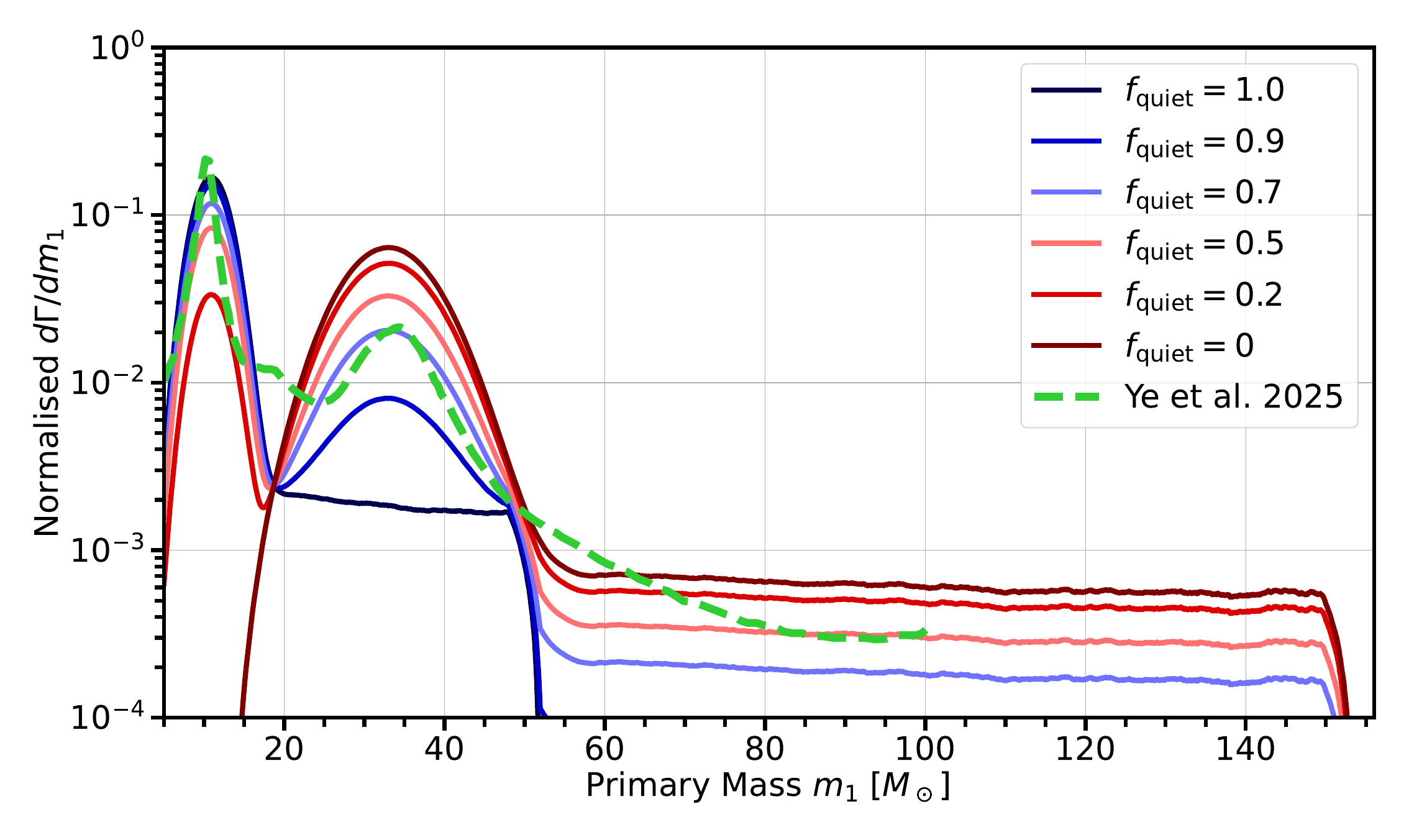}
    \includegraphics[width=0.49\textwidth]
    {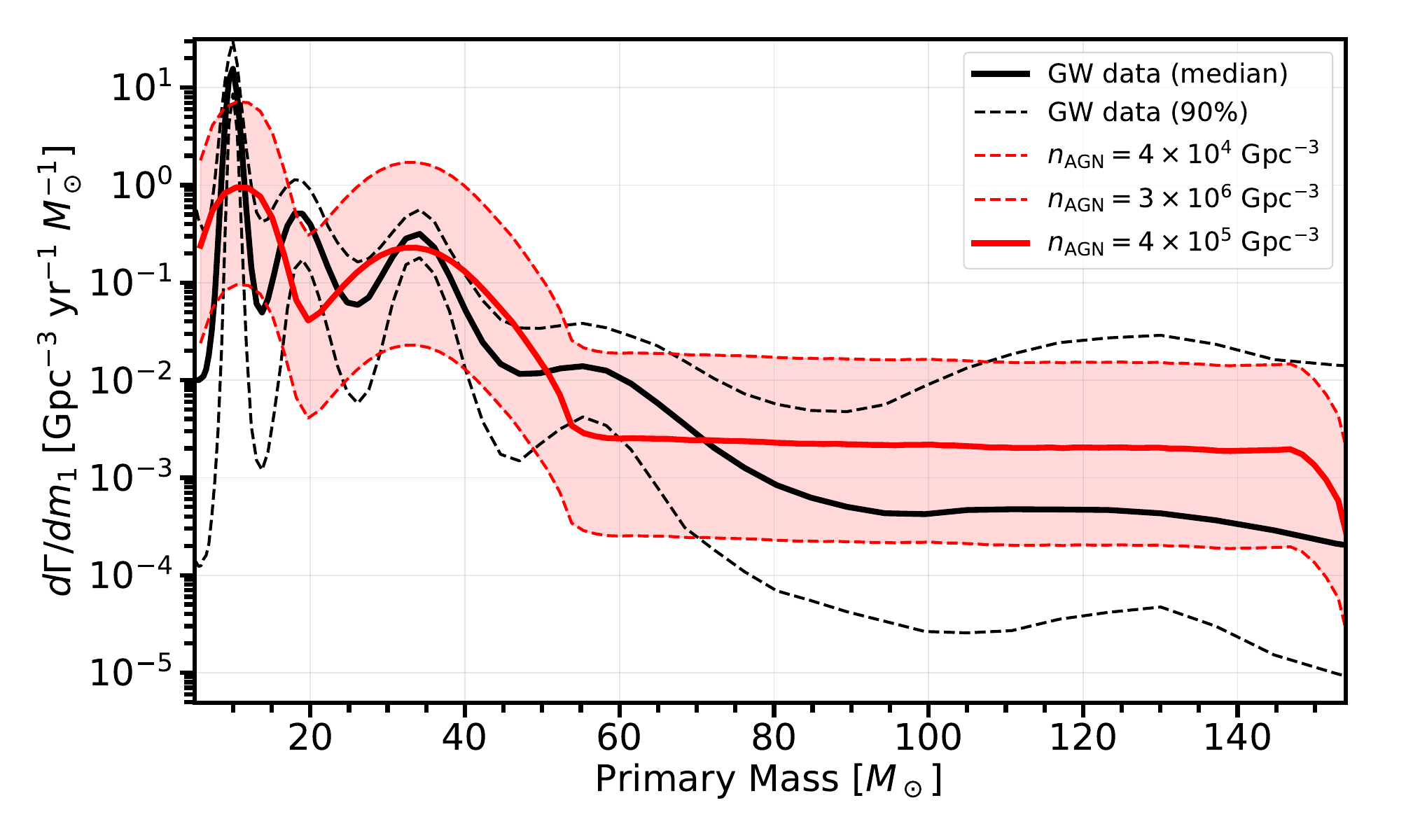}
    \caption{
Probability distribution of the gravitational-wave merger rate as a function of primary mass. The distributions are derived from the evolving compact-object populations shown in previous sections and therefore represent the observable consequences of accretion-driven transport through mass space. Top Left: evolution of a fully accreting population, illustrating the progressive development of a high-mass merger tail. Top Right: population consisting of 70\% non-accreting and 30\% accreting binaries. Bottom Left: merger-rate distributions after $50\,{\rm{Myr}}$ for different fractions of non-accreting systems, $f_{\rm quiet}$. Mixed populations naturally produce multi-component merger distributions whose relative amplitudes trace the fraction of binaries embedded in long-lived gas reservoirs.
Lower Right: Merger rate per primary mass as derived from our model after $50 \rm{Myr}$, assuming that $70\%$ of the population are non-accreting binaries, and the rest are accreting. The shaded region describes different choices of $n_{\rm{AGN}}$ in the range $4\times 10^4-3\times 10^6 \ \rm{Gpc}^{-3}$ compared to the LVK-inferred rate; the 90th percentile is shown as dashed lines \citep[as adopted from][]{Ginat2026}.
}
    \label{fig:dGamma}
\end{figure*}

Figure~\ref{fig:dGamma} shows the evolution of the merger-rate distribution as a function of primary mass.  For comparison, Figure~\ref{fig:dGamma} also shows the phenomenological merger-mass distribution based on LVK data \citep[adopted from][]{Ginat2026}. The top panel illustrates the evolution of the mass distribution over time when all the binaries accrete. 
As accretion proceeds, the merger population extends to progressively larger masses, reflecting the transport of compact objects through mass space discussed in Section~\ref{sec:distributions}. The high-mass tail of the merger distribution is therefore not imposed by the initial conditions, but emerges naturally as compact objects are driven to larger masses through sustained gas accretion.

In reality, not all compact-object binaries are expected to experience identical accretion histories and conditions. All the panels, apart from the top-left one, therefore, explore mixed populations consisting of both accreting and non-accreting systems. The top-left one corresponds to a population in which all the binaries accrete. The resulting merger distributions exhibit multiple components, with one peak associated with the unprocessed population and another associated with binaries that have undergone substantial gas-driven growth. The relative prominence of these features is controlled by the fraction of systems embedded in long-lived gas reservoirs.

These results suggest that gravitational-wave catalogs may already contain observable signatures of accretion-driven transport. In addition to extending the merger population toward higher masses, mixtures of accreting and non-accreting binaries can generate multi-component mass distributions that retain information about the fraction of systems that experienced significant gas-driven growth. 
Events such as GW231123 may represent the high-mass extreme of a broader population shaped by sustained accretion-driven growth.

\section{Discussion}\label{sec:discussion}

\subsection{Accretion-Driven Transport Across Gaseous Environments}

Although our numerical examples were motivated by AGN disks, the framework developed here is more general. The central result of this work is that gas accretion transports compact objects through mass space, with the resulting population evolution determined primarily by the mass dependence of the accretion rate. Consequently, the qualitative behaviour described here is expected to arise in a broad range of gas-rich environments capable of sustaining compact-object accretion.

AGN disks provide a particularly compelling example because they combine large gas reservoirs with long dynamical lifetimes and are expected to host substantial populations of compact objects, making them a fertile ground for gravitational wave mergers \citep{Bartos2017,Stone2017,McKernan2018,Tagawa2020}. However, they are unlikely to be unique. Other dense stellar systems may also undergo phases of substantial gas retention. Observations of multiple stellar populations in globular clusters have motivated scenarios in which clusters experience secondary episodes of star formation within dense gaseous environments \citep[see the review by][]{BastianLardo2018}. If there are compact objects embedded during such phases, gas accretion can alter their masses and binary properties in a manner analogous to that discussed here, in the relevant accretion regimes. Gas will also affect the binary evolution leading to merger in these environments \citep{RoznerPerets2023}.

More recently, gas-rich proto-clusters and embedded stellar systems have been proposed as environments in which stellar-mass black holes can experience substantial accretion-driven growth \citep{Roupas2025}. Collective accretion may be enhanced in such systems because the surrounding cluster potential modifies the gas flow and increases the amount of material available to individual accretors \citep{LinMurray2007,2019ApJ...876..142K}.
It was suggested
that gas-assisted growth may contribute to the formation of black holes in the pair-instability mass gap and potentially even intermediate-mass black holes.

While the detailed gas properties differ substantially among these environments, the continuity-equation framework developed in Section~\ref{sec:distributions} suggests that their influence on compact-object populations can be understood through the accretion law. Environments characterized by $\beta>1$ should naturally generate broadened mass distributions and extended high-mass tails, whereas environments with $\beta<1$ should tend to compress the population toward a narrower range of masses. The masses, mass ratios, and merger properties of compact-object binaries may therefore retain some information about the gaseous environments through which they evolved.

From this perspective, gaseous environments should not be viewed solely as locations where compact-object mergers occur.
Rather, they act as transport media that continuously reshape compact-object populations through accretion, imprinting the properties of their environments onto the gravitational-wave sources they ultimately produce.

\subsection{Spin Evolution During Gas Accretion}

Throughout this work, we have focused on the evolution of masses and mass ratios. However, gas accretion is also expected to modify other binary properties, such as the spin distribution of compact objects. In the simplest picture, the spin evolution of an accreting black hole is determined by the angular momentum carried by the accreted gas together with the increase in the black-hole mass.

For a black hole of mass $m$ and angular momentum $J$, the dimensionless spin parameter is
\begin{equation}
a_\ast=\frac{cJ}{Gm^2}.
\end{equation}
If the accreted gas carries specific angular momentum $j_{\rm acc}$, then
\begin{equation}
\dot J = \dot m j_{\rm acc},
\end{equation}
and the spin evolves according to
\begin{equation}
\dot a_\ast
=
\frac{c}{Gm^2}\dot J
-
2a_\ast\frac{\dot m}{m}
=
\frac{\dot m}{m}
\left(
\frac{c j_{\rm acc}}{Gm}
-
2a_\ast
\right).
\end{equation}

In the absence of additional torques, accretion therefore produces a flow not only through mass space but also through spin space. A joint distribution $p(m,a_\ast,t)$ would obey a continuity equation analogous to that developed in Section~\ref{sec:distributions},
\begin{equation}
\frac{\partial p}{\partial t}
+
\frac{\partial}{\partial m}
\left(
\dot m\,p
\right)
+
\frac{\partial}{\partial a_\ast}
\left(
\dot a_\ast\,p
\right)
=0.
\end{equation}

In this simple picture, coherent accretion naturally spins black holes up as they grow \citep{Bardeen1970,Thorne1974}, suggesting that the same transport framework used to describe mass evolution could, in principle, be extended to predict spin distributions.

Modern GRMHD simulations have shown, however, that the situation is considerably more complex. Large-scale magnetic fields threading the event horizon can extract angular momentum and rotational energy through the Blandford--Znajek mechanism \citep{Blandford1977}. In magnetically arrested disks (MADs), the electromagnetic torques associated with relativistic jets can rival or exceed the angular momentum supplied by the accreting gas, producing efficient spin-down despite continued mass growth \citep{Tchekhovskoy2011,Tchekhovskoy2012,Narayan2022,Lowell2025}. 
Recent MAD simulations suggest that electromagnetic spin extraction can balance or even exceed the angular momentum supplied by the accreting gas, leading to equilibrium spins substantially below the near-extremal values predicted by classical thin-disk theory.

The spin distribution of compact objects in gaseous environments, therefore, depends not only on the amount of accreted mass but also on the magnetic state and geometry of the accretion flow. While the transport framework developed here provides a predictive description of mass-function evolution, extending it to spins requires further understanding of magnetic-flux accumulation, jet production, and electromagnetic angular-momentum extraction. High-mass mergers produced through gas-assisted growth may therefore exhibit spin distributions that depend as much on the magnetic state of their accretion environment as on the total amount of mass they accreted.

\subsection{Caveats and Future Directions}

The framework presented here isolates the effects of accretion-driven transport in mass space and demonstrates how the mass dependence of the accretion rate can reshape compact-object populations. Several physical processes may modify the quantitative predictions while leaving the qualitative picture unchanged.

First, our calculations employ idealized prescriptions for gas accretion, focusing on the Bondi--Hoyle--Lyttleton, Hill, and Eddington-limited regimes. While the normalization of the growth rates depends on the details of the accretion flow, the emergence of high-mass tails is primarily a consequence of differential growth and is therefore expected to persist whenever the accretion rate increases with mass. Future work should explore more sophisticated accretion prescriptions calibrated by hydrodynamic and magnetohydrodynamic simulations.

Second, we have neglected feedback from the accreting compact objects. Radiation, outflows, and jets may modify the local gas supply and alter the efficiency of accretion. Such effects are likely to regulate the overall growth rate and may influence the transition between different accretion regimes. Incorporating feedback self-consistently remains an important challenge for future population models.

Third, we have focused on the evolution of pre-existing binaries. In gaseous environments, however, gas dynamical interactions can also facilitate the formation of new binaries and higher-order multiples. These systems may subsequently undergo accretion-driven growth and mergers, potentially enhancing the merger rates predicted here. Our treatment should therefore be regarded as conservative in the sense that it neglects additional channels through which gas can assemble and catalyse merging compact-object systems.

Finally, we have adopted a simplified description of the gaseous environment. We assume circular binaries embedded in a uniform medium characterized by fixed local conditions throughout their evolution. In realistic systems, binaries may migrate, evolve through eccentric or inclined configurations, and experience time-dependent gas properties associated with environmental evolution or duty cycles \citep[e.g.,][]{Su2025}. Such effects will modify the accretion history of individual systems and may lead to populations that are best described as superpositions of multiple transport histories, weighted according to the different dynamical scenarios.

These caveats highlight the need for future studies that combine accretion-driven transport with self-consistent models of binary dynamics, feedback, migration, and environmental evolution. Nevertheless, the results presented here suggest that the qualitative consequences of differential accretion are robust: gas-rich environments can systematically reshape compact-object populations by transporting objects through mass space, leaving observable signatures in the gravitational-wave sources they ultimately produce.

\section{Summary}\label{sec:summary}

In this \textit{Letter}, we developed a general framework for understanding how gas accretion reshapes compact-object populations. We showed that the evolution of an accreting population can be viewed as a transport problem in mass space, governed by a continuity equation and determined primarily by the mass dependence of the accretion rate laws. Accretion laws with $\beta>1$ produce divergent evolution and naturally generate extended high-mass tails, while accretion laws with $\beta<1$ lead to convergent evolution and compress the population toward a narrower range of masses.

We connected this general framework to physically motivated accretion prescriptions, including Bondi--Hoyle--Lyttleton, Hill, and Eddington-limited growth, and explored their consequences using both analytical calculations and Monte Carlo population models. We showed that gas accretion can substantially broaden compact-object mass distributions and populate regions of parameter space that are difficult to access through stellar evolution alone. In particular, accretion-driven growth naturally produces a population of high-mass black holes and offers a potential pathway toward explaining massive gravitational-wave events such as GW231123.

We further demonstrated that binaries provide an additional diagnostic through the evolution of their mass ratios. When binary components accrete independently, differential growth tends to deepen initial mass asymmetries. In contrast, once the accretion radii overlap and a common wake develops, enhanced accretion onto the secondary drives the population toward equal masses. The resulting mass-ratio distribution therefore depends sensitively on how the captured gas is partitioned between the binary components.

Finally, we showed that these population-level changes leave observable signatures in gravitational-wave merger populations. Accretion-driven transport produces extended high-mass merger tails and, in mixed populations containing both accreting and non-accreting systems, naturally generates multi-component merger distributions. While our numerical examples were motivated by AGN disks, the underlying framework applies more broadly to compact objects evolving within gaseous environments. From this perspective, gas-rich environments are not merely locations where mergers occur; they act as transport media that continuously reshape compact-object populations through accretion, leaving observable imprints on the gravitational-wave sources they ultimately produce.

\section*{Acknowledgments}

MR gratefully acknowledges support from the Institute for Advanced Study, the Kovner Member Fund, and the Gruber Foundation Fellowship. The UCSC team is supported in part by the Heising-Simons Foundation, the Vera Rubin Presidential Chair for Diversity at UCSC, and the National Science Foundation (AST-2307710, AST-2206243, AST-1911206). This research was also made possible through funding from the Lamat Institute \citep{2025NatAs...9.1770Q}.

\bibliography{sample701}{}

@ARTICLE{2025NatAs...9.1770Q,
       author = {{Quinteros}, Katherine N. and {Covarrubias}, Rebecca and {Ramirez-Ruiz}, Enrico},
        title = "{Mobilizing the strengths of marginalized students in STEM research programmes}",
      journal = {Nature Astronomy},
         year = 2025,
        month = dec,
       volume = {9},
        pages = {1770-1775},
          doi = {10.1038/s41550-025-02736-y},
       adsurl = {https://ui.adsabs.harvard.edu/abs/2025NatAs...9.1770Q}
}

@ARTICLE{Lowell2025,
       author = {{Lowell}, Beverly H. and {Tchekhovskoy}, Alexander and {Gottlieb}, Ore and {Natarajan}, Priyamvada},
        title = "{A semi-analytic model for black hole spin evolution in magnetically arrested disks}",
      journal = {\mnras},
         year = 2023,
        month = nov,
       volume = {526},
       number = {3},
        pages = {4291-4308},
          doi = {10.1093/mnras/stad2993},
       adsurl = {https://ui.adsabs.harvard.edu/abs/2023MNRAS.526.4291L}
}

@article{Thorne1974,
  author = {Thorne, Kip S.},
  title = {Disk-Accretion onto a Black Hole. II. Evolution of the Hole},
  journal = {Astrophysical Journal},
  volume = {191},
  pages = {507--520},
  year = {1974},
  doi = {10.1086/152991}
}

@article{Blandford1977,
  author = {Blandford, Roger D. and Znajek, Roman L.},
  title = {Electromagnetic Extraction of Energy from Kerr Black Holes},
  journal = {Monthly Notices of the Royal Astronomical Society},
  volume = {179},
  pages = {433--456},
  year = {1977},
  doi = {10.1093/mnras/179.3.433}
}

@article{Tchekhovskoy2011,
  author = {Tchekhovskoy, Alexander and Narayan, Ramesh and McKinney, Jonathan C.},
  title = {Efficient Generation of Jets from Magnetically Arrested Accretion on a Rapidly Spinning Black Hole},
  journal = {Monthly Notices of the Royal Astronomical Society: Letters},
  volume = {418},
  pages = {L79--L83},
  year = {2011},
  doi = {10.1111/j.1745-3933.2011.01147.x}
}

@article{Tchekhovskoy2012,
  author = {Tchekhovskoy, Alexander and McKinney, Jonathan C. and Narayan, Ramesh},
  title = {General Relativistic Modeling of Magnetically Choked Accretion Flows and Jets},
  journal = {Journal of Physics: Conference Series},
  volume = {372},
  pages = {012040},
  year = {2012},
  doi = {10.1088/1742-6596/372/1/012040}
}

@article{Narayan2022,
  author = {Narayan, Ramesh and Chael, Andrew and Chatterjee, Koushik and Ricarte, Angelo and Curd, Brandon},
  title = {Jets in Magnetically Arrested Hot Accretion Flows: Geometry, Power, and Black Hole Spin-Down},
  journal = {Monthly Notices of the Royal Astronomical Society},
  volume = {511},
  pages = {3795--3813},
  year = {2022},
  doi = {10.1093/mnras/stac285}
}

@article{Bardeen1970,
  author = {Bardeen, James M.},
  title = {Kerr Metric Black Holes},
  journal = {Nature},
  volume = {226},
  pages = {64--65},
  year = {1970},
  doi = {10.1038/226064a0}
}

@ARTICLE{2019ApJ...876..142K,
       author = {{Kaaz}, Nicholas and {Antoni}, Andrea and {Ramirez-Ruiz}, Enrico},
        title = "{Bondi-Hoyle-Lyttleton Accretion onto Star Clusters}",
      journal = {\apj},
         year = 2019,
        month = may,
       volume = {876},
       number = {2},
          eid = {142},
        pages = {142},
          doi = {10.3847/1538-4357/ab158b},
archivePrefix = {arXiv},
       eprint = {1901.03649},
 primaryClass = {astro-ph.HE},
       adsurl = {https://ui.adsabs.harvard.edu/abs/2019ApJ...876..142K}
}

@ARTICLE{2023ApJ...944...44K,
       author = {{Kaaz}, Nicholas and {Schr{\o}der}, Sophie Lund and {Andrews}, Jeff J. and {Antoni}, Andrea and {Ramirez-Ruiz}, Enrico},
        title = "{The Hydrodynamic Evolution of Binary Black Holes Embedded within the Vertically Stratified Disks of Active Galactic Nuclei}",
      journal = {\apj},
         year = 2023,
        month = feb,
       volume = {944},
       number = {1},
          eid = {44},
        pages = {44},
          doi = {10.3847/1538-4357/aca967},
archivePrefix = {arXiv},
       eprint = {2103.12088},
 primaryClass = {astro-ph.HE},
       adsurl = {https://ui.adsabs.harvard.edu/abs/2023ApJ...944...44K}
}

@ARTICLE{FirstGW,
       author = {{Abbott}, B.~P. and {Abbott}, R. and {Abbott}, T.~D. and {Abernathy}, M.~R. and {Acernese}, F. and {Ackley}, K. and {Adams}, C. and {Adams}, T. and {Addesso}, P. and {Adhikari}, R.~X. and {Adya}, V.~B. and {Affeldt}, C. and {Agathos}, M. and {Agatsuma}, K. and {Aggarwal}, N. and {Aguiar}, O.~D. and {Aiello}, L. and {Ain}, A. and {Ajith}, P. and {Allen}, B. and {Allocca}, A. and {Altin}, P.~A. and {Anderson}, S.~B. and {Anderson}, W.~G. and {Arai}, K. and {Arain}, M.~A. and {Araya}, M.~C. and {Arceneaux}, C.~C. and {Areeda}, J.~S. and {Arnaud}, N. and {Arun}, K.~G. and {Ascenzi}, S. and {Ashton}, G. and {Ast}, M. and {Aston}, S.~M. and {Astone}, P. and {Aufmuth}, P. and {Aulbert}, C. and {Babak}, S. and {Bacon}, P. and {Bader}, M.~K.~M. and {Baker}, P.~T. and {Baldaccini}, F. and {Ballardin}, G. and {Ballmer}, S.~W. and {Barayoga}, J.~C. and {Barclay}, S.~E. and {Barish}, B.~C. and {Barker}, D. and {Barone}, F. and {Barr}, B. and {Barsotti}, L. and {Barsuglia}, M. and {Barta}, D. and {Bartlett}, J. and {Barton}, M.~A. and {Bartos}, I. and {Bassiri}, R. and {Basti}, A. and {Batch}, J.~C. and {Baune}, C. and {Bavigadda}, V. and {Bazzan}, M. and {Behnke}, B. and {Bejger}, M. and {Belczynski}, C. and {Bell}, A.~S. and {Bell}, C.~J. and {Berger}, B.~K. and {Bergman}, J. and {Bergmann}, G. and {Berry}, C.~P.~L. and {Bersanetti}, D. and {Bertolini}, A. and {Betzwieser}, J. and {Bhagwat}, S. and {Bhandare}, R. and {Bilenko}, I.~A. and {Billingsley}, G. and {Birch}, J. and {Birney}, I.~A. and {Birnholtz}, O. and {Biscans}, S. and {Bisht}, A. and {Bitossi}, M. and {Biwer}, C. and {Bizouard}, M.~A. and {Blackburn}, J.~K. and {Blair}, C.~D. and {Blair}, D.~G. and {Blair}, R.~M. and {Bloemen}, S. and {Bock}, O. and {Bodiya}, T.~P. and {Boer}, M. and {Bogaert}, G. and {Bogan}, C. and {Bohe}, A. and {Bojtos}, P. and {Bond}, C. and {Bondu}, F. and {Bonnand}, R. and {Boom}, B.~A. and {Bork}, R. and {Boschi}, V. and {Bose}, S. and {Bouffanais}, Y. and {Bozzi}, A. and {Bradaschia}, C. and {Brady}, P.~R. and {Braginsky}, V.~B. and {Branchesi}, M. and {Brau}, J.~E. and {Briant}, T. and {Brillet}, A. and {Brinkmann}, M. and {Brisson}, V. and {Brockill}, P. and {Brooks}, A.~F. and {Brown}, D.~A. and {Brown}, D.~D. and {Brown}, N.~M. and {Buchanan}, C.~C. and {Buikema}, A. and {Bulik}, T. and {Bulten}, H.~J. and {Buonanno}, A. and {Buskulic}, D. and {Buy}, C. and {Byer}, R.~L. and {Cabero}, M. and {Cadonati}, L. and {Cagnoli}, G. and {Cahillane}, C. and {Bustillo}, J. Calder{\'o}n and {Callister}, T. and {Calloni}, E. and {Camp}, J.~B. and {Cannon}, K.~C. and {Cao}, J. and {Capano}, C.~D. and {Capocasa}, E. and {Carbognani}, F. and {Caride}, S. and {Diaz}, J. Casanueva and {Casentini}, C. and {Caudill}, S. and {Cavagli{\`a}}, M. and {Cavalier}, F. and {Cavalieri}, R. and {Cella}, G. and {Cepeda}, C.~B. and {Baiardi}, L. Cerboni and {Cerretani}, G. and {Cesarini}, E. and {Chakraborty}, R. and {Chalermsongsak}, T. and {Chamberlin}, S.~J. and {Chan}, M. and {Chao}, S. and {Charlton}, P. and {Chassande-Mottin}, E. and {Chen}, H.~Y. and {Chen}, Y. and {Cheng}, C. and {Chincarini}, A. and {Chiummo}, A. and {Cho}, H.~S. and {Cho}, M. and {Chow}, J.~H. and {Christensen}, N. and {Chu}, Q. and {Chua}, S. and {Chung}, S. and {Ciani}, G. and {Clara}, F. and {Clark}, J.~A. and {Cleva}, F. and {Coccia}, E. and {Cohadon}, P. -F. and {Colla}, A. and {Collette}, C.~G. and {Cominsky}, L. and {Constancio}, M. and {Conte}, A. and {Conti}, L. and {Cook}, D. and {Corbitt}, T.~R. and {Cornish}, N. and {Corsi}, A. and {Cortese}, S. and {Costa}, C.~A. and {Coughlin}, M.~W. and {Coughlin}, S.~B. and {Coulon}, J. -P. and {Countryman}, S.~T. and {Couvares}, P. and {Cowan}, E.~E. and {Coward}, D.~M. and {Cowart}, M.~J.},
        title = "{Observation of Gravitational Waves from a Binary Black Hole Merger}",
      journal = {\prl},
         year = 2016,
        month = feb,
       volume = {116},
       number = {6},
          eid = {061102},
        pages = {061102},
          doi = {10.1103/PhysRevLett.116.061102},
archivePrefix = {arXiv},
       eprint = {1602.03837},
 primaryClass = {gr-qc},
       adsurl = {https://ui.adsabs.harvard.edu/abs/2016PhRvL.116f1102A}
}

@ARTICLE{GW231123detection,
       author = {{The LIGO Scientific Collaboration} and {the Virgo Collaboration} and {the KAGRA Collaboration} and {Abac}, A.~G. and {Abouelfettouh}, I. and {Acernese}, F. and {Ackley}, K. and {Adamcewicz}, C. and {Adhicary}, S. and {Adhikari}, D. and {Adhikari}, N. and {Adhikari}, R.~X. and {Adkins}, V.~K. and {Afroz}, S. and {Agapito}, A. and {Agarwal}, D. and {Agathos}, M. and {Aggarwal}, N. and {Aggarwal}, S. and {Aguiar}, O.~D. and {Ahrend}, I. -L. and {Aiello}, L. and {Ain}, A. and {Ajith}, P. and {Akutsu}, T. and {Albanesi}, S. and {Ali}, W. and {Al-Kershi}, S. and {All{\'e}n{\'e}}, C. and {Allocca}, A. and {Al-Shammari}, S. and {Altin}, P.~A. and {Alvarez-Lopez}, S. and {Amar}, W. and {Amarasinghe}, O. and {Amato}, A. and {Amicucci}, F. and {Amra}, C. and {Ananyeva}, A. and {Anderson}, S.~B. and {Anderson}, W.~G. and {Andia}, M. and {Ando}, M. and {Andr{\'e}s-Carcasona}, M. and {Andri{\'c}}, T. and {Anglin}, J. and {Ansoldi}, S. and {Antelis}, J.~M. and {Antier}, S. and {Aoumi}, M. and {Appavuravther}, E.~Z. and {Appert}, S. and {Apple}, S.~K. and {Arai}, K. and {Araujo Alvarez}, C. and {Araya}, A. and {Araya}, M.~C. and {Arca Sedda}, M. and {Areeda}, J.~S. and {Aritomi}, N. and {Armato}, F. and {Armstrong}, S. and {Arnaud}, N. and {Arogeti}, M. and {Aronson}, S.~M. and {Arun}, K.~G. and {Ashton}, G. and {Aso}, Y. and {Asprea}, L. and {Assiduo}, M. and {Assis de Souza Melo}, S. and {Aston}, S.~M. and {Astone}, P. and {Attadio}, F. and {Aubin}, F. and {AultONeal}, K. and {Avallone}, G. and {Avila}, E.~A. and {Babak}, S. and {Badger}, C. and {Bae}, S. and {Bagnasco}, S. and {Baiotti}, L. and {Bajpai}, R. and {Baka}, T. and {Baker}, A.~M. and {Baker}, K.~A. and {Baker}, T. and {Baldi}, G. and {Baldicchi}, N. and {Ball}, M. and {Ballardin}, G. and {Ballmer}, S.~W. and {Banagiri}, S. and {Banerjee}, B. and {Bankar}, D. and {Baptiste}, T.~M. and {Baral}, P. and {Baratti}, M. and {Barayoga}, J.~C. and {Barish}, B.~C. and {Barker}, D. and {Barman}, N. and {Barneo}, P. and {Barone}, F. and {Barr}, B. and {Barsotti}, L. and {Barsuglia}, M. and {Barta}, D. and {Bartoletti}, A.~M. and {Barton}, M.~A. and {Bartos}, I. and {Basalaev}, A. and {Bassiri}, R. and {Basti}, A. and {Bawaj}, M. and {Baxi}, P. and {Bayley}, J.~C. and {Baylor}, A.~C. and {Baynard}, II, P.~A. and {Bazzan}, M. and {Bedakihale}, V.~M. and {Beirnaert}, F. and {Bejger}, M. and {Belardinelli}, D. and {Bell}, A.~S. and {Bellie}, D.~S. and {Bellizzi}, L. and {Benoit}, W. and {Bentara}, I. and {Bentley}, J.~D. and {Ben Yaala}, M. and {Bera}, S. and {Bergamin}, F. and {Berger}, B.~K. and {Bernuzzi}, S. and {Beroiz}, M. and {Berry}, C.~P.~L. and {Bersanetti}, D. and {Bertheas}, T. and {Bertolini}, A. and {Betzwieser}, J. and {Beveridge}, D. and {Bevilacqua}, G. and {Bevins}, N. and {Bhandare}, R. and {Bhatt}, R. and {Bhattacharjee}, D. and {Bhattacharyya}, S. and {Bhaumik}, S. and {Bhagwat}, S. and {Biancalana}, V. and {Bianchi}, A. and {Bilenko}, I.~A. and {Billingsley}, G. and {Binetti}, A. and {Bini}, S. and {Binu}, C. and {Biot}, S. and {Birnholtz}, O. and {Biscoveanu}, S. and {Bisht}, A. and {Bitossi}, M. and {Bizouard}, M. -A. and {Blaber}, S. and {Blackburn}, J.~K. and {Blagg}, L.~A. and {Blair}, C.~D. and {Blair}, D.~G. and {Bode}, N. and {Boettner}, N. and {Boileau}, G. and {Boldrini}, M. and {Bolingbroke}, G.~N. and {Bolliand}, A. and {Bonavena}, L.~D. and {Bondarescu}, R. and {Bondu}, F. and {Bonilla}, E. and {Bonilla}, M.~S. and {Bonino}, A. and {Bonnand}, R. and {Borchers}, A. and {Borhanian}, S. and {Boschi}, V. and {Bose}, S. and {Bossilkov}, V. and {Bothra}, Y. and {Boudon}, A. and {Bourg}, L. and {Bouyer}, G. and {Boyle}, M. and {Bozzi}, A. and {Bradaschia}, C. and {Brady}, P.~R. and {Branch}, A. and {Branchesi}, M. and {Braun}, I. and {Briant}, T. and {Brillet}, A.},
        title = "{GW231123: a Binary Black Hole Merger with Total Mass 190-265 $M_{\odot}$}",
      journal = {arXiv e-prints},
         year = 2025,
        month = jul,
          eid = {arXiv:2507.08219},
        pages = {arXiv:2507.08219},
          doi = {10.48550/arXiv.2507.08219},
archivePrefix = {arXiv},
       eprint = {2507.08219},
 primaryClass = {astro-ph.HE},
       adsurl = {https://ui.adsabs.harvard.edu/abs/2025arXiv250708219T}
}

@ARTICLE{Antoni2019,
       author = {{Antoni}, Andrea and {MacLeod}, Morgan and {Ramirez-Ruiz}, Enrico},
        title = "{The Evolution of Binaries in a Gaseous Medium: Three-dimensional Simulations of Binary Bondi-Hoyle-Lyttleton Accretion}",
      journal = {\apj},
         year = 2019,
        month = oct,
       volume = {884},
       number = {1},
          eid = {22},
        pages = {22},
          doi = {10.3847/1538-4357/ab3466},
archivePrefix = {arXiv},
       eprint = {1901.07572},
 primaryClass = {astro-ph.HE},
       adsurl = {https://ui.adsabs.harvard.edu/abs/2019ApJ...884...22A}
}

@ARTICLE{Antonini2025,
       author = {{Antonini}, Fabio and {Romero-Shaw}, Isobel M. and {Callister}, Thomas},
        title = "{Star Cluster Population of High Mass Black Hole Mergers in Gravitational Wave Data}",
      journal = {Physical Review Letters},
         year = 2025,
        month = jan,
       volume = {134},
       number = {1},
          eid = {011401},
        pages = {011401},
          doi = {10.1103/PhysRevLett.134.011401},
archivePrefix = {arXiv},
       eprint = {2406.19044},
 primaryClass = {astro-ph.HE},
       adsurl = {https://ui.adsabs.harvard.edu/abs/2025PhRvL.134a1401A}
}

@ARTICLE{Abramowicz1988,
       author = {{Abramowicz}, M.~A. and {Czerny}, B. and {Lasota}, J.~P. and {Szuszkiewicz}, E.},
        title = "{Slim Accretion Disks}",
      journal = {\apj},
         year = 1988,
        month = sep,
       volume = {332},
        pages = {646},
          doi = {10.1086/166683},
       adsurl = {https://ui.adsabs.harvard.edu/abs/1988ApJ...332..646A}
}

@ARTICLE{Bartos2017,
       author = {{Bartos}, Imre and {Kocsis}, Bence and {Haiman}, Zolt{\'a}n and {M{\'a}rka}, Szabolcs},
        title = "{Rapid and Bright Stellar-mass Binary Black Hole Mergers in Active Galactic Nuclei}",
      journal = {\apj},
         year = 2017,
        month = feb,
       volume = {835},
       number = {2},
          eid = {165},
        pages = {165},
          doi = {10.3847/1538-4357/835/2/165},
archivePrefix = {arXiv},
       eprint = {1602.03831},
 primaryClass = {astro-ph.HE},
       adsurl = {https://ui.adsabs.harvard.edu/abs/2017ApJ...835..165B}
}

@ARTICLE{BartosHaiman2025,
       author = {{Bartos}, Imre and {Haiman}, Zoltan},
        title = "{Accretion is All You Need: Black Hole Spin Alignment in Merger GW231123 Indicates Accretion Pathway}",
      journal = {arXiv e-prints},
         year = 2025,
        month = aug,
          eid = {arXiv:2508.08558},
        pages = {arXiv:2508.08558},
          doi = {10.48550/arXiv.2508.08558},
archivePrefix = {arXiv},
       eprint = {2508.08558},
 primaryClass = {astro-ph.HE},
       adsurl = {https://ui.adsabs.harvard.edu/abs/2025arXiv250808558B}
}

@ARTICLE{BastianLardo2018,
       author = {{Bastian}, Nate and {Lardo}, Carmela},
        title = "{Multiple Stellar Populations in Globular Clusters}",
      journal = {\araa},
         year = 2018,
        month = sep,
       volume = {56},
        pages = {83-136},
          doi = {10.1146/annurev-astro-081817-051839},
archivePrefix = {arXiv},
       eprint = {1712.01286},
 primaryClass = {astro-ph.SR},
       adsurl = {https://ui.adsabs.harvard.edu/abs/2018ARA&A..56...83B}
}

@ARTICLE{Bondi1952,
       author = {{Bondi}, H.},
        title = "{On spherically symmetrical accretion}",
      journal = {\mnras},
         year = 1952,
        month = jan,
       volume = {112},
        pages = {195},
          doi = {10.1093/mnras/112.2.195},
       adsurl = {https://ui.adsabs.harvard.edu/abs/1952MNRAS.112..195B}
}

@ARTICLE{Bonnell2001a,
       author = {{Bonnell}, I.~A. and {Bate}, M.~R. and {Clarke}, C.~J. and {Pringle}, J.~E.},
        title = "{Competitive accretion in embedded stellar clusters}",
      journal = {\mnras},
         year = 2001,
        month = may,
       volume = {323},
       number = {4},
        pages = {785-794},
          doi = {10.1046/j.1365-8711.2001.04270.x},
archivePrefix = {arXiv},
       eprint = {astro-ph/0102074},
 primaryClass = {astro-ph},
       adsurl = {https://ui.adsabs.harvard.edu/abs/2001MNRAS.323..785B}
}

@ARTICLE{Bonnell2001b,
       author = {{Bonnell}, I.~A. and {Clarke}, C.~J. and {Bate}, M.~R. and {Pringle}, J.~E.},
        title = "{Accretion in stellar clusters and the initial mass function}",
      journal = {\mnras},
         year = 2001,
        month = jun,
       volume = {324},
       number = {3},
        pages = {573-579},
          doi = {10.1046/j.1365-8711.2001.04311.x},
archivePrefix = {arXiv},
       eprint = {astro-ph/0102121},
 primaryClass = {astro-ph},
       adsurl = {https://ui.adsabs.harvard.edu/abs/2001MNRAS.324..573B}
}

@ARTICLE{Comerford2019,
       author = {{Comerford}, T.~A.~F. and {Izzard}, R.~G. and {Booth}, R.~A. and {Rosotti}, G.},
        title = "{Bondi-Hoyle-Lyttleton accretion by binary stars}",
      journal = {\mnras},
         year = 2019,
        month = dec,
       volume = {490},
       number = {4},
        pages = {5196-5209},
          doi = {10.1093/mnras/stz2977},
archivePrefix = {arXiv},
       eprint = {1910.13353},
 primaryClass = {astro-ph.SR},
       adsurl = {https://ui.adsabs.harvard.edu/abs/2019MNRAS.490.5196C}
}

@ARTICLE{Fishbach2017,
       author = {{Fishbach}, Maya and {Holz}, Daniel E. and {Farr}, Ben},
        title = "{Are LIGO's Black Holes Made from Smaller Black Holes?}",
      journal = {\apjl},
         year = 2017,
        month = may,
       volume = {840},
       number = {2},
          eid = {L24},
        pages = {L24},
          doi = {10.3847/2041-8213/aa7045},
archivePrefix = {arXiv},
       eprint = {1703.06869},
 primaryClass = {astro-ph.HE},
       adsurl = {https://ui.adsabs.harvard.edu/abs/2017ApJ...840L..24F}
}

@ARTICLE{Gerosa2015,
       author = {{Gerosa}, D. and {Veronesi}, B. and {Lodato}, G. and {Rosotti}, G.},
        title = "{Spin alignment and differential accretion in merging black hole binaries}",
      journal = {\mnras},
         year = 2015,
        month = aug,
       volume = {451},
       number = {4},
        pages = {3941-3954},
          doi = {10.1093/mnras/stv1214},
archivePrefix = {arXiv},
       eprint = {1503.06807},
 primaryClass = {astro-ph.GA},
       adsurl = {https://ui.adsabs.harvard.edu/abs/2015MNRAS.451.3941G}
}

@ARTICLE{Grobner2020,
       author = {{Gr{\"o}bner}, M. and {Ishibashi}, W. and {Tiwari}, S. and {Haney}, M. and {Jetzer}, P.},
        title = "{Binary black hole mergers in AGN accretion discs: gravitational wave rate density estimates}",
      journal = {\aap},
         year = 2020,
        month = jun,
       volume = {638},
          eid = {A119},
        pages = {A119},
          doi = {10.1051/0004-6361/202037681},
archivePrefix = {arXiv},
       eprint = {2005.03571},
 primaryClass = {astro-ph.GA},
       adsurl = {https://ui.adsabs.harvard.edu/abs/2020A&A...638A.119G}
}

@ARTICLE{HoyleLyttleton1941,
       author = {{Hoyle}, F. and {Lyttleton}, R.~A.},
        title = "{On the accretion theory of stellar evolution}",
      journal = {\mnras},
         year = 1941,
        month = jan,
       volume = {101},
        pages = {227},
          doi = {10.1093/mnras/101.4.227},
       adsurl = {https://ui.adsabs.harvard.edu/abs/1941MNRAS.101..227H}
}

@ARTICLE{Kiroglu2025,
       author = {{K{\i}ro{\u{g}}lu}, Fulya and {Kremer}, Kyle and {Rasio}, Frederic A.},
        title = "{Beyond Hierarchical Mergers: Accretion-Driven Origins of Massive, Highly Spinning Black Holes in Dense Star Clusters}",
      journal = {arXiv e-prints},
         year = 2025,
        month = sep,
          eid = {arXiv:2509.05415},
        pages = {arXiv:2509.05415},
          doi = {10.48550/arXiv.2509.05415},
archivePrefix = {arXiv},
       eprint = {2509.05415},
 primaryClass = {astro-ph.HE},
       adsurl = {https://ui.adsabs.harvard.edu/abs/2025arXiv250905415K}
}

@ARTICLE{LinMurray2007,
       author = {{Lin}, Douglas N.~C. and {Murray}, Stephen D.},
        title = "{Gas Accretion by Globular Clusters and Nucleated Dwarf Galaxies and the Formation of the Arches and Quintuplet Clusters}",
      journal = {\apj},
         year = 2007,
        month = jun,
       volume = {661},
       number = {2},
        pages = {779-786},
          doi = {10.1086/515387},
archivePrefix = {arXiv},
       eprint = {astro-ph/0703807},
 primaryClass = {astro-ph},
       adsurl = {https://ui.adsabs.harvard.edu/abs/2007ApJ...661..779L}
}

@ARTICLE{McKernan2012,
       author = {{McKernan}, B. and {Ford}, K.~E.~S. and {Lyra}, W. and {Perets}, H.~B.},
        title = "{Intermediate mass black holes in AGN discs - I. Production and growth}",
      journal = {\mnras},
         year = 2012,
        month = sep,
       volume = {425},
       number = {1},
        pages = {460-469},
          doi = {10.1111/j.1365-2966.2012.21486.x},
archivePrefix = {arXiv},
       eprint = {1206.2309},
 primaryClass = {astro-ph.GA},
       adsurl = {https://ui.adsabs.harvard.edu/abs/2012MNRAS.425..460M}
}

@ARTICLE{McKernan2018,
       author = {{McKernan}, Barry and {Ford}, K.~E. Saavik and {Bellovary}, J. and {Leigh}, N.~W.~C. and {Haiman}, Z. and {Kocsis}, B. and {Lyra}, W. and {Mac Low}, M.-M. and {Metzger}, B. and {O'Dowd}, M. and {Endlich}, S. and {Rosen}, D.~J.},
        title = "{Constraining Stellar-mass Black Hole Mergers in AGN Disks Detectable with LIGO}",
      journal = {\apj},
         year = 2018,
        month = oct,
       volume = {866},
       number = {1},
          eid = {66},
        pages = {66},
          doi = {10.3847/1538-4357/aadae5},
archivePrefix = {arXiv},
       eprint = {1702.07818},
 primaryClass = {astro-ph.HE},
       adsurl = {https://ui.adsabs.harvard.edu/abs/2018ApJ...866...66M}
}

@ARTICLE{Patton2022,
       author = {{Patton}, Rachel A. and {Sukhbold}, Tuguldur and {Eldridge}, J.~J.},
        title = "{Comparing compact object distributions from mass- and presupernova core structure-based prescriptions}",
      journal = {\mnras},
         year = 2022,
        month = mar,
       volume = {511},
       number = {1},
        pages = {903-913},
          doi = {10.1093/mnras/stab3797},
archivePrefix = {arXiv},
       eprint = {2106.05978},
 primaryClass = {astro-ph.HE},
       adsurl = {https://ui.adsabs.harvard.edu/abs/2022MNRAS.511..903P}
}

@ARTICLE{Peters1964,
       author = {{Peters}, P.~C.},
        title = "{Gravitational Radiation and the Motion of Two Point Masses}",
      journal = {Physical Review},
         year = 1964,
        month = nov,
       volume = {136},
       number = {4B},
        pages = {1224-1232},
          doi = {10.1103/PhysRev.136.B1224},
       adsurl = {https://ui.adsabs.harvard.edu/abs/1964PhRv..136.1224P}
}

@article{RakavyShaviv1967,
  author       = {Rakavy, G. and Shaviv, G.},
  title        = {Instabilities in Highly Evolved Stellar Models},
  journal      = {The Astrophysical Journal},
  volume       = {148},
  pages        = {803},
  year         = {1967},
  doi          = {10.1086/149204},
  publisher    = {University of Chicago Press},
}

@ARTICLE{Rosselli-Calderon2026,
       author = {{Rosselli-Calderon}, Alejandra and {Stewart}, Julia and {Shen}, Sijing and {Chakrabarti}, Sukanya and {Soares-Furtado}, Melinda and {Ramirez-Ruiz}, Enrico},
        title = "{Chemical Enrichment of Metal-poor Stars Orbiting Massive Black Hole Companions}",
      journal = {\apj},
         year = 2026,
        month = jan,
       volume = {997},
       number = {1},
          eid = {13},
        pages = {13},
          doi = {10.3847/1538-4357/ae28c6},
archivePrefix = {arXiv},
       eprint = {2508.14163},
 primaryClass = {astro-ph.SR},
       adsurl = {https://ui.adsabs.harvard.edu/abs/2026ApJ...997...13R}
}

@ARTICLE{Roupas2025,
       author = {{Roupas}, Zacharias},
        title = "{Black hole mass function shift in proto-stellar clusters driven by gas accretion}",
      journal = {\aap},
         year = 2025,
        month = oct,
       volume = {702},
          eid = {A208},
        pages = {A208},
          doi = {10.1051/0004-6361/202556434},
archivePrefix = {arXiv},
       eprint = {2509.08448},
 primaryClass = {astro-ph.GA},
       adsurl = {https://ui.adsabs.harvard.edu/abs/2025A&A...702A.208R}
}

@ARTICLE{RoznerPerets2023,
       author = {{Rozner}, Mor and {Perets}, Hagai B.},
        title = "{Binary Evolution, Gravitational-wave Mergers, and Explosive Transients in Multiple-population Gas-enriched Globular Clusters}",
      journal = {\apj},
         year = 2022,
        month = jun,
       volume = {931},
       number = {2},
          eid = {149},
        pages = {149},
          doi = {10.3847/1538-4357/ac6d55},
archivePrefix = {arXiv},
       eprint = {2203.01330},
 primaryClass = {astro-ph.HE},
       adsurl = {https://ui.adsabs.harvard.edu/abs/2022ApJ...931..149R}
}

@ARTICLE{RoznerPerets2024,
       author = {{Rozner}, Mor and {Perets}, Hagai B.},
        title = "{Soft No More: Gas Shielding Protects Soft Binaries from Disruption in Gas-rich Environments}",
      journal = {\apj},
         year = 2024,
        month = jun,
       volume = {968},
       number = {2},
          eid = {80},
        pages = {80},
          doi = {10.3847/1538-4357/ad4bdd},
archivePrefix = {arXiv},
       eprint = {2404.01384},
 primaryClass = {astro-ph.HE},
       adsurl = {https://ui.adsabs.harvard.edu/abs/2024ApJ...968...80R}
}

@ARTICLE{Salpeter1955,
       author = {{Salpeter}, Edwin E.},
        title = "{The Luminosity Function and Stellar Evolution.}",
      journal = {\apj},
         year = 1955,
        month = jan,
       volume = {121},
        pages = {161},
          doi = {10.1086/145971},
       adsurl = {https://ui.adsabs.harvard.edu/abs/1955ApJ...121..161S}
}

@ARTICLE{Stone2017,
       author = {{Stone}, Nicholas C. and {Metzger}, Brian D. and {Haiman}, Zolt{\'a}n},
        title = "{Assisted inspirals of stellar mass black holes embedded in AGN discs: solving the `final au problem'}",
      journal = {\mnras},
         year = 2017,
        month = jan,
       volume = {464},
       number = {1},
        pages = {946-954},
          doi = {10.1093/mnras/stw2260},
archivePrefix = {arXiv},
       eprint = {1602.04226},
 primaryClass = {astro-ph.GA},
       adsurl = {https://ui.adsabs.harvard.edu/abs/2017MNRAS.464..946S}
}

@ARTICLE{Su2025,
       author = {{Su}, Yubo and {Rowan}, Connar and {Rozner}, Mor},
        title = "{Gas meets Kozai: the influence of a gas-rich accretion disc on hierarchical triples undergoing von Zeipel{\textendash}Lidov{\textendash}Kozai oscillations}",
      journal = {\mnras},
         year = 2025,
        month = oct,
       volume = {543},
       number = {2},
        pages = {1864-1877},
          doi = {10.1093/mnras/staf1592},
archivePrefix = {arXiv},
       eprint = {2505.23889},
 primaryClass = {astro-ph.GA},
       adsurl = {https://ui.adsabs.harvard.edu/abs/2025MNRAS.543.1864S}
}

@ARTICLE{Sukhbold2016,
       author = {{Sukhbold}, Tuguldur and {Ertl}, T. and {Woosley}, S.~E. and {Brown}, Justin M. and {Janka}, H. -T.},
        title = "{Core-collapse Supernovae from 9 to 120 Solar Masses Based on Neutrino-powered Explosions}",
      journal = {\apj},
         year = 2016,
        month = apr,
       volume = {821},
       number = {1},
          eid = {38},
        pages = {38},
          doi = {10.3847/0004-637X/821/1/38},
archivePrefix = {arXiv},
       eprint = {1510.04643},
 primaryClass = {astro-ph.HE},
       adsurl = {https://ui.adsabs.harvard.edu/abs/2016ApJ...821...38S}
}

@ARTICLE{Tagawa2020,
       author = {{Tagawa}, Hiromichi and {Haiman}, Zolt{\'a}n and {Kocsis}, Bence},
        title = "{Formation and Evolution of Compact-object Binaries in AGN Disks}",
      journal = {\apj},
         year = 2020,
        month = jul,
       volume = {898},
       number = {1},
          eid = {25},
        pages = {25},
          doi = {10.3847/1538-4357/ab9b8c},
archivePrefix = {arXiv},
       eprint = {1912.08218},
 primaryClass = {astro-ph.GA},
       adsurl = {https://ui.adsabs.harvard.edu/abs/2020ApJ...898...25T}
}

@ARTICLE{WoosleyHeger2021,
       author = {{Woosley}, S.~E. and {Heger}, Alexander},
        title = "{The Pair-instability Mass Gap for Black Holes}",
      journal = {\apjl},
         year = 2021,
        month = may,
       volume = {912},
       number = {2},
          eid = {L31},
        pages = {L31},
          doi = {10.3847/2041-8213/abf2c4},
archivePrefix = {arXiv},
       eprint = {2103.07933},
 primaryClass = {astro-ph.SR},
       adsurl = {https://ui.adsabs.harvard.edu/abs/2021ApJ...912L..31W}
}

@ARTICLE{Yang2019a,
       author = {{Yang}, Y. and {Bartos}, I. and {Haiman}, Z. and {Kocsis}, B. and {M{\'a}rka}, Z. and {Stone}, N.~C. and {M{\'a}rka}, S.},
        title = "{AGN Disks Harden the Mass Distribution of Stellar-mass Binary Black Hole Mergers}",
      journal = {\apj},
         year = 2019,
        month = may,
       volume = {876},
       number = {2},
          eid = {122},
        pages = {122},
          doi = {10.3847/1538-4357/ab16e3},
archivePrefix = {arXiv},
       eprint = {1903.01405},
 primaryClass = {astro-ph.HE},
       adsurl = {https://ui.adsabs.harvard.edu/abs/2019ApJ...876..122Y}
}

@ARTICLE{Yang2019,
       author = {{Yang}, Y. and {Bartos}, I. and {Gayathri}, V. and {Ford}, K.~E.~S. and {Haiman}, Z. and {Klimenko}, S. and {Kocsis}, B. and {M{\'a}rka}, S. and {M{\'a}rka}, Z. and {McKernan}, B. and {O'Shaughnessy}, R.},
        title = "{Hierarchical Black Hole Mergers in Active Galactic Nuclei}",
      journal = {\prl},
         year = 2019,
        month = nov,
       volume = {123},
       number = {18},
          eid = {181101},
        pages = {181101},
          doi = {10.1103/PhysRevLett.123.181101},
archivePrefix = {arXiv},
       eprint = {1906.09281},
 primaryClass = {astro-ph.HE},
       adsurl = {https://ui.adsabs.harvard.edu/abs/2019PhRvL.123r1101Y}
}

@ARTICLE{YoungClarke2015,
       author = {{Young}, M.~D. and {Clarke}, C.~J.},
        title = "{Binary accretion rates: dependence on temperature and mass ratio}",
      journal = {\mnras},
         year = 2015,
        month = sep,
       volume = {452},
       number = {3},
        pages = {3085-3091},
          doi = {10.1093/mnras/stv1512},
archivePrefix = {arXiv},
       eprint = {1507.01850},
 primaryClass = {astro-ph.SR},
       adsurl = {https://ui.adsabs.harvard.edu/abs/2015MNRAS.452.3085Y}
}

@ARTICLE{Zinnecker1982,
       author = {{Zinnecker}, H.},
        title = "{Prediction of the protostellar mass spectrum in the Orion near-infrared cluster}",
      journal = {Annals of the New York Academy of Sciences},
         year = 1982,
        month = oct,
       volume = {395},
        pages = {226-235},
          doi = {10.1111/j.1749-6632.1982.tb43399.x},
       adsurl = {https://ui.adsabs.harvard.edu/abs/1982NYASA.395..226Z}
}

@ARTICLE{Tagawa2021,
       author = {{Tagawa}, Hiromichi and {Kocsis}, Bence and {Haiman}, Zolt{\'a}n and {Bartos}, Imre and {Omukai}, Kazuyuki and {Samsing}, Johan},
        title = "{Mass-gap Mergers in Active Galactic Nuclei}",
      journal = {\apj},
         year = 2021,
        month = feb,
       volume = {908},
       number = {2},
          eid = {194},
        pages = {194},
          doi = {10.3847/1538-4357/abd555},
archivePrefix = {arXiv},
       eprint = {2012.00011},
 primaryClass = {astro-ph.HE},
       adsurl = {https://ui.adsabs.harvard.edu/abs/2021ApJ...908..194T}
}

@ARTICLE{Tagawa2026,
       author = {{Tagawa}, Hiromichi and {Haiman}, Zolt{\'a}n and {Kocsis}, Bence},
        title = "{Properties of black hole mergers in disks of active galactic nuclei}",
      journal = {arXiv e-prints},
         year = 2026,
        month = apr,
          eid = {arXiv:2604.25994},
        pages = {arXiv:2604.25994},
          doi = {10.48550/arXiv.2604.25994},
archivePrefix = {arXiv},
       eprint = {2604.25994},
 primaryClass = {astro-ph.HE},
       adsurl = {https://ui.adsabs.harvard.edu/abs/2026arXiv260425994T}
}

@ARTICLE{Ginat2026,
       author = {{Ginat}, Yonadav Barry and {Antonini}, Fabio and {Flanagan}, Elizabeth and {Gieles}, Mark},
        title = "{Second-Generation Mass Peak in the Gravitational-Wave Population as a Probe of Globular Clusters}",
      journal = {arXiv e-prints},
         year = 2026,
        month = apr,
          eid = {arXiv:2604.07456},
        pages = {arXiv:2604.07456},
          doi = {10.48550/arXiv.2604.07456},
archivePrefix = {arXiv},
       eprint = {2604.07456},
 primaryClass = {astro-ph.HE},
       adsurl = {https://ui.adsabs.harvard.edu/abs/2026arXiv260407456G}
}

@ARTICLE{Opik1924,
       author = {{{\"O}pik}, E.},
        title = "{Statistical Studies of Double Stars: On the Distribution of Relative Luminosities and Distances of Double Stars in the Harvard Revised Photometry North of Declination -31{\textdegree}}",
      journal = {Publications of the Tartu Astrofizica Observatory},
         year = 1924,
        month = jan,
       volume = {25},
        pages = {1},
       adsurl = {https://ui.adsabs.harvard.edu/abs/1924PTarO..25f...1O}
}
\bibliographystyle{aasjournalv7}

\end{document}